\documentclass[preprint,12pt]{elsarticle}

\usepackage{amssymb}
\usepackage{amsmath}
\usepackage{booktabs}
\usepackage{array}
\usepackage{xcolor}
\usepackage{graphicx}
\usepackage{multirow}
\usepackage{xpatch}
\usepackage{float}
\usepackage{colortbl}
\definecolor{lightgray}{gray}{0.95}
\definecolor{lightred}{rgb}{1.0, 0.85, 0.85}
\definecolor{lightgreen}{rgb}{0.85, 1.0, 0.85}
\usepackage{color,soul}
\usepackage{makecell}

\makeatletter
\renewcommand\paragraph{\@startsection{paragraph}{4}{\z@}%
    {-3.25ex \@plus -1ex \@minus -.2ex}%
    {-1em}%
    {\normalfont\normalsize\itshape}}
\makeatother

\begin{document}

\begin{frontmatter}



\title{Adaptive Voxel-Weighted Loss Using L1 Norms in Deep Neural Networks for Detection and Segmentation of Prostate
Cancer Lesions in PET/CT Images}

\renewcommand{\thefootnote}{\fnsymbol{footnote}}
\author[a,b,c]{Obed K. Dzikunu\footnotemark[1]}
\author[a,b]{Shadab Ahamed\footnotemark[2]}
\author[a,b]{Amirhossein Toosi\footnotemark[2]}
\author[b]{Sara Harsini}
\author[a,b]{Fran\c{c}ois B\a'{e}nard}
\author[a,c]{Xiaoxiao Li}
\author[a,b]{Arman Rahmim}


\affiliation[a]{organization={University of British Columbia},
            city={Vancouver},
            country={Canada}}
            
\affiliation[b]{organization={BC Cancer Research Institute},
            city={Vancouver},
            country={Canada}}

\affiliation[c]{organization={Vector Institute},
            city={Toronto},
            country={Canada}}



\begin{abstract}

\textbf{Background and Objective:} Accurate automated detection of recurrent prostate cancer in PSMA PET/CT scans is challenging due to heterogeneous lesion size, activity, anatomical location, and intra- and inter-class imbalances. Conventional deep learning loss functions often produce suboptimal optimization, as gradients are dominated by easy background voxels or extreme outliers. To address this, we propose L1-weighted Dice Focal Loss (L1DFL), which harmonizes gradient magnitudes across voxels using L1 norms to adaptively weight samples based on classification difficulty, resulting in well-calibrated predictions with a bimodal separation between correct and incorrect predictions.

\textbf{Methods:} We trained three 3D convolutional networks (Attention U-Net, SegResNet, U-Net) and a transformer-based UNETR model on 380 PSMA PET/CT scans. PET and CT volumes were concatenated as input. A finetuned medical foundation model (SAM-Med3D) was also evaluated. L1DFL was compared against Dice Loss (DL) and Dice Focal Loss (DFL) using metrics including Dice Similarity Coefficient (DSC), F1 score, and lesion-wise detection performance, with statistical significance assessed via Wilcoxon signed-rank tests with Bonferroni correction ($\alpha = 0.003$).

\textbf{Results:} Across architectures, L1DFL consistently outperformed DL and DFL, achieving at least a $4\%$ improvement in DSC. F1 scores were higher by $\ge6\%$ and $\ge26\%$ compared to DL and DFL, respectively. While DFL produced more false positives and DL struggled with larger lesions, L1DFL achieved balanced detection, minimizing false detections while maintaining high true positive rates. The gradient harmonization mechanism ensured robustness across varying lesion sizes, volumes, and spread.

\textbf{Conclusion:} L1DFL improves segmentation of recurrent prostate cancer lesions in PSMA PET/CT by harmonizing gradients. It produces well-calibrated predictions with clear separation of correct and incorrect cases, and balanced detection across lesion sizes while maintaining consistent performance across multiple architectures.

\end{abstract}


\begin{highlights}
\item Besides uptake, tumor characteristics such as volume and dissemination limit automated detection performance.
\item The proposed loss function demonstrates strong calibration, producing a clear separation between correct and incorrect predictions.
\item It dynamically distributes gradient magnitudes across samples, avoiding overemphasis on outliers or overly easy examples.
\item It outperforms the Dice and Dice Focal Losses by at least $4\%$ based on Dice Score.
\item It outperforms Dice Loss and Dice Focal Losses by $6\%$ and $26\%$, respectively, on F1 score.
\end{highlights}

\renewcommand{\thefootnote}{\fnsymbol{footnote}}
\footnotetext[1]{Corresponding author: Obed Korshie Dzikunu (email: okdzikunu@gmail.com)}
\footnotetext[2]{These authors contributed equally to this work.}

\end{frontmatter}



\section{Introduction}
\label{intro}
Semantic segmentation assigns a label to every pixel or voxel in an image \cite{farabet2012learning}, requiring both local feature extraction and broader contextual understanding \cite{chen2014semantic}, since spatial coherence and boundary smoothness are critical \cite{7478072}. In medical imaging, segmentation accuracy is particularly critical, as false detections or missed lesions can significantly affect diagnosis, staging, and treatment planning \cite{macmanus2009use}. These difficulties are compounded by severe class imbalance, where pathological voxels constitute only a small fraction of the image volume.

In whole-body oncological PET, lesion voxels are heavily outnumbered by background, and lesions vary widely in size, shape, contrast, and tracer uptake \cite{hofheinz2013automatic}. Metastatic prostate cancer (mPCa) amplifies these challenges further, with lesions distributed across multiple anatomical sites \cite{fendler201768}. PSMA PET imaging adds another layer of complexity, as PSMA expression appears in normal organs and inflammatory processes as well as malignant tissue \cite{lauri2022psma}. AI-based detection systems must therefore balance sensitivity to small or low-uptake lesions against specificity to avoid false positives from physiological uptake.

Most automated approaches struggle with this trade-off. Techniques such as standardized uptake value (SUV) thresholding are commonly employed post hoc to suppress false positives by classifying voxels above a predefined threshold as foreground \cite{foster2014review}. While effective in reducing spurious detections, such strategies often sacrifice sensitivity for lesions with low tracer avidity or lesions in patients with atypical uptake patterns \cite{li2024automated}. More sophisticated methods have attacked the problem at the input or representation level: semi-automatic pipelines with bone masking and SUV thresholding \cite{gafita2019qpsma}, nnU-Net-based whole-body frameworks \cite{jafari2024convolutional,kendrick2022fully}, anatomical organ priors \cite{li2024automated}, 2D diffusion models on MIP projections \cite{toosi2024segment}, and hybrid MIP-based frameworks \cite{constantino2025use}. Feature-clipping normalization \cite{bhandary2024segmentation} and 2.5D pipelines \cite{zhao2020deep} have also been explored. While each has yielded incremental gains, false positives and missed small or ambiguous lesions remain persistent failure modes, and critically, none directly addresses how voxel-level errors are weighted during optimization.

In PET lesion segmentation, not all voxels contribute equally to learning. The majority of background voxels are trivially classified, while a small subset of foreground and boundary voxels is consistently misclassified or uncertain. Standard losses do not distinguish between these difficulty levels. Region-based losses such as Dice Loss (DL) handle inter-class imbalance through overlap-based formulation \cite{10.1007/978-3-319-46976-8_19,Zhang2021DiceLoss} but treat all voxels within a class equally \cite{9338261}. Distribution-based losses like Focal Loss \cite{lin2017focal} and TopK loss \cite{wu2016bridging} address difficulty-based weighting but rely on fixed hyperparameters that limit adaptability to mini-batch distributions \cite{10.1007/978-3-030-32226-7_10}, and improper weighting can compromise sensitivity-specificity balance \cite{YEUNG2022102026}. Compound losses such as Dice Focal Loss \cite{zhu2019anatomynet} and Combo Loss \cite{taghanaki2019combo} leverage complementary strengths but inherit the same static weighting limitations.

A more principled solution is gradient harmonization \cite{li2019gradient}, which adaptively weights samples based on gradient density rather than fixed hyperparameters, preventing easy examples from overwhelming the contribution of hard minority cases. Liu et al. \cite{10.1007/978-3-030-32226-7_10} extended this to medical image segmentation via knee MRI, demonstrating its value for handling intra-class imbalance. However, this approach has not yet been explored for PET lesion segmentation, where the gradient density problem is especially acute.

We therefore propose \textit{L1-weighted Dice Focal Loss} (L1DFL), which integrates gradient harmonization into the Dice Focal Loss framework. Unlike prior approaches that bin samples by computed gradient magnitudes \cite{li2019gradient,10.1007/978-3-030-32226-7_10}, L1DFL uses the L1 norm between predicted probabilities and ground truth labels as a canonical proxy for voxel-level classification difficulty \cite{levin2017markov,peyre2019computational,LIU2024103015}. Voxels in high-density difficulty regions are down-weighted while those in low-density regions are up-weighted, distributing gradient contributions uniformly across the difficulty spectrum and avoiding domination by any single difficulty level during optimization.

The main contributions of this work are: (i) a novel loss function, L1DFL, incorporating gradient harmonization; (ii) evaluation of L1DFL against DL and DFL on false positive and false negative rates; (iii) comparison across single-lesion and multiple-lesion scenarios; (iv) assessment on clinical metrics including molecular tumor volume and lesion spread; and (v) benchmarking across five models: Attention U-Net \cite{oktay2018attention}, SegResNet \cite{myronenko20193d}, U-Net \cite{cciccek20163d}, UNETR \cite{hatamizadeh2022unetr}, and SAM-Med3D \cite{wang2024sam}.

\section{Materials and Methods}
\label{method}
\subsection{Dataset and Ground Truth Annotation}

We analyzed 380 [$^{18}$F]DCFPyL PET/CT scans of patients diagnosed with biochemical recurrence of prostate cancer as part of an ongoing clinical trial (NCT02899312) with informed written consent obtained from the participants. The image data collection was approved by the University of British Columbia – BC Cancer Ethics Board. The inclusion criteria for the patients were: (1) histologically proven prostate cancer with biochemical recurrence after initial curative therapy with radical prostatectomy, with a PSA $>$ 0.4 ng/mL and an additional PSA measurement confirming an increase; and (2) histologically proven prostate cancer with biochemical recurrence after initial curative radiotherapy, with a PSA $>$ 2 ng/mL after therapy, with five or fewer lesions identified on [$^{18}$F]DCFPyL PET/CT \cite{harsini2023outcome}.

Patients received an average of 350 MBq of [$^{18}$F]DCFPyL following a four-hour fast, with dose adjustments based on body weight. PET/CT imaging was performed 120 minutes post-injection using GE Discovery 600 or 690 scanners (GE Healthcare, USA). A non-contrast CT scan was acquired for attenuation correction and anatomical localization (120 kV, automated mA modulation), followed by a whole-body PET acquisition lasting 2-4 minutes per bed position. PET images were reconstructed using ordered subset expectation maximization with point-spread function modeling, producing transaxial images with a matrix resolution of $192 \times 192$ and an in-plane pixel spacing of $3.64~\text{mm}$ x $3.64~\text{mm}^2$ and a slice thickness of $3.27~\text{mm}$ \cite{harsini2023outcome}.

There were a total of 684 prostate cancer (PCa) lesions across the entire dataset with a mean active lesion volume of 6.68 $\pm$ 10.20 ml. The lesions varied in uptake with the average SUVmax and mean standardized uptake value (SUVmean) being 12.65$\pm$14.46 and 4.62 $\pm$ 3.88, respectively. These were segmented by a board-certified nuclear medicine physician from the BC Cancer Research Institute to obtain segmentation masks serving as ground truth. All lesions were annotated using the PET Edge tool, a semi-automated gradient-based segmentation tool, and contours refined using the 3D Brush tool. Both tools are available in the MIM workstation (MIM software, Ohio, USA).

\subsection{Data Preprocessing and Augmentation}

After image acquisition, PET activity values were converted to decay-corrected standardized uptake values (SUV), while CT intensities (in Hounsfield units) were clipped to the range $[-1000, 3000]$ and normalized to $[0, 1]$ using min-max scaling. PET images were left in their absolute SUV values. All images, including PET, CT, and ground truth (GT) masks, were resampled to an isotropic voxel spacing of $[2 \text{ mm}, 2 \text{ mm}, 2 \text{ mm}]$ using bilinear interpolation for PET and CT, and nearest-neighbor interpolation for GT masks. For data augmentation, training images underwent random cropping and affine transformations, including translations in (-10,10) voxels along each spatial dimension, rotations up to $\frac{\pi}{15}$ around the z axis, and isotropic scaling sampled uniformly from 0.9 to 1.1. Cubic patches of size $128 \times 128 \times 128$ voxels were extracted, with an $80\%$ probability of being centered on a foreground voxel. The augmented CT and PET patches were concatenated along the channel dimension and used as input to the networks, with lesion detection primarily driven by PET and CT providing anatomical context. For SAM-Med3D, a uni-modal model, only PET images were used as input.

\subsection{Implementation Details}

\paragraph{Architectures:} We evaluated five volumetric segmentation architectures: SegResNet \cite{myronenko20193d}, Attention U-Net \cite{oktay2018attention}, 3D U-Net \cite{cciccek20163d}, UNETR \cite{hatamizadeh2022unetr}, and SAM-Med3D \cite{wang2024sam}. SegResNet, Attention U-Net, and 3D U-Net were implemented as convolutional encoder--decoder networks with multi-channel inputs and two-class outputs. SegResNet employed encoder depths of \{1, 2, 2, 4\} with an initial channel size of 16, doubling at each downsampling stage, and used Group Normalization and ReLU activations. Attention U-Net consisted of five encoder levels with channel sizes \{16, 32, 64, 128, 256\}, Batch Normalization, and attention gates as originally proposed. The 3D U-Net comprised six levels with channels increasing from 16 to 512, residual blocks with Batch Normalization, and a symmetric decoder with transposed convolutions.

UNETR and SAM-Med3D represent transformer-based and foundation-model approaches. UNETR used a vision transformer encoder with hidden size 256, MLP dimension 1024, and four attention heads, convolutional patch embeddings, and a U-shaped convolutional decoder. SAM-Med3D is a fully three-dimensional extension of the Segment Anything Model designed for volumetric medical image segmentation. It follows the original SAM design, consisting of a 3D image encoder, a 3D prompt encoder, and a mask decoder, with all 2D operations replaced by 3D counterparts, including 3D convolutions, normalization layers, and positional encodings \cite{wang2024sam}. We used the publicly released SAM-Med3D pretrained weights and fine-tuned the model using our proposed and comparative loss functions.

\paragraph{Model Training:} We divided the dataset into training, validation, and testing sets containing 258, 65, and 57 samples, respectively. The training objective was to minimize the loss functions in the training set. We used AdamW optimizer with a weight decay of $10^{-5}$ to optimize the loss functions. We adopted a cosine annealing scheduler to decay the learning rate from $2 \times 10^{-4}$ to zero for 1000 epochs. The loss was first computed for each batch within an epoch, and the overall loss for an epoch was calculated by taking a mean over all the batch losses. The model with the highest mean DSC in the validation phase was chosen for further evaluation of the test set.

For SAM-Med3D, we evaluated both zero-shot performance and fine-tuning. Fine-tuning was conducted using the original loss function as well as comparative loss functions for evaluation. Training used a batch size of 2, an initial learning rate of $8\times10^{-6}$ and 150 epochs. Optimization employed AdamW with a linear learning rate scheduler (decay factor 0.1).

\subsection{Loss Functions}
\paragraph{Dice Loss:}
The DL maximizes the overlap between the predicted segmentation and the ground truth. In this work, for a specific cubic patch of an image, the DL for a batch was computed as,

\begin{equation}
\label{eq:DiceLoss}
   \mathcal{L}_\text{Dice} = 1 - \frac{1}{2}\sum_{c \in \{0, 1\}} \frac{2 \sum_i p_i (c) g_i(c)}{\sum_i p_i(c) + \sum_i g_i (c) + \epsilon}
\end{equation}
where \( p_i \) and \( g_i \) are, respectively, the predicted probability and ground truth for a given class \( c \) of the voxel \( i \), and \( \epsilon \) is a small constant to prevent division by zero.

\paragraph{Focal Loss:}
The Focal Loss \cite{lin2017focal} helps focus on hard-to-classify examples by down-weighting the easy ones. In this work, the Focal Loss for an image patch was computed over a batch as,

\begin{equation}
\label{eq:Focalloss}
    \mathcal{L}_{\text{Focal}} = - \sum_{c \in \{0, 1\}}\sum_i g_i(c)\alpha_c (1 - p_i(c))^\gamma \log(p_i(c))
\end{equation}
where \( \alpha \) is a factor defining the balance between background and foreground classes in a binary segmentation task, and \( \gamma \) is the focusing parameter that adjusts the rate at which easy examples are down-weighted.

\paragraph{Dice Focal Loss:}
The DFL \cite{zhu2019anatomynet} combines the DL and the Focal Loss forming a compound loss function:
\begin{equation}
    \mathcal{L}_{\text{DFL}} = \mathcal{L}_{\text{Dice}} + \mathcal{L}_{\text{Focal}}
\end{equation}
For the DFL formulation, we set the Focal Loss terms \( \gamma =2 \) and  \( \alpha  = 1\) in all our experiments similar to the implementation reported in \cite{10.1007/978-3-031-09002-8_9}.

\paragraph{L1-weighted Dice Focal Loss:}
L1DFL applies a gradient harmonization weighting strategy to the Dice Loss based on both the L1 norm between predicted probabilities and ground truth labels (quantifying voxel-level difficulty) and the density of samples at each difficulty level (defined by bins), then combines this weighted Dice Loss with Focal Loss. The DL generally has two variants, one with squared terms in the denominator \cite{milletari2016v} and one without \cite{10.1007/978-3-319-46976-8_19}. In L1DFL, we employ the DL with the squared denominator which was preferred over the non-squared version following its performance in \cite{10.1007/978-3-030-32226-7_10}. We describe the mathematical formulation of the weighting strategy next.

First, we compute the L1 norm, \(\Delta\), between the predicted probabilities \( p \) and the ground truth labels \( g \),
\begin{equation}
\Delta_i = \| g_i - p_i \|_1, \text{ for \(i = 1,2, ... N\) } \label{eq:L1norm}
\end{equation}
where N is the total number of samples.
Next, we partition the range of L1 norm values \(\Delta\) into bins of a consistent nominal bin width of \(\Gamma\). Each bin \(B\) is defined by its center \(B_k\):

\begin{equation}
B_k = \Gamma\cdot k, \text{for \(k = 0,1,2, ... n-1\)} \label{eq:bin}
\end{equation}
where \(n=\lceil{\frac{1}{\Gamma}}+1\rceil\) is the total number of bins. For instance, with a bin width of \(\Gamma = 0.1\) (\(n\) = 11 bins), \(B_k\) takes values at regular intervals in the range [0,1], such that for \(k = 0, 1, 2, \dots, 10\), bin centers (\(B_k\)) = \([0, 0.1, 0.2, \dots, 1.0]\).

We then calculate the count of \(\Delta\) values that fall within each bin:
\begin{equation}
\mathcal{C}(B_k) = \sum_{i=1}^N \delta_\kappa(B_k, \Delta_i) \label{eq:bincount}
\end{equation}
where \(\delta_\kappa(B_k, \Delta_i)\) is an indicator function defined as:
\begin{equation}
\delta_\kappa(B_k, \Delta_i) = 
\begin{cases}
1, & \text{if } |B_k - \Delta_i| \leq \frac{\Gamma}{2} \\
0, & \text{otherwise}
\end{cases} \label{eq:indicator}
\end{equation}
To account for potential truncation near the boundaries of [0,1], we calculate the effective bin width for each bin as:
\begin{equation}
\lambda_\kappa(B_k) = \min \Big\{B_k + \frac{\Gamma}{2}, 1\Big\} - \max\Big\{B_k - \frac{\Gamma}{2}, 0\Big\} \label{eq:binWidth}
\end{equation}
Using \(\lambda_\kappa(B_k)\), the norm density, \(\mathcal{D}(B)\), for each bin is defined as:
\begin{equation}
\mathcal{D}(B_k) = \frac{C(B_k)}{\lambda_\kappa(B_k)} \label{eq:density}
\end{equation}

The density $\mathcal{D}(B_k)$ quantifies how many voxels have L1 norms within the bin centered at $B_k$. The bin center represents the difficulty level: $B_k \approx 0$ indicates easy samples (predictions closely match ground truth), while $B_k \approx 1$ indicates hard samples (predictions deviate significantly from ground truth). Thus, density, $\mathcal{D}(B_k)$, measures the prevalence of each difficulty level, and high density means many voxels share similar L1 norm values (commonly occurring difficulty), while low density means few voxels fall in that range. The weight for each bin is thus calculated as:

\begin{equation}
w(B_k) = \frac{N}{\mathcal{D}(B_k)} \label{eq:weight}
\end{equation}

where $N$ is the total number of samples. Consequently, samples in high-density regions receive lower weights, while samples in low-density regions receive higher weights. This ensures that less frequent and difficult samples, which often correspond to small foreground structures in imbalanced scenarios, receive appropriate emphasis without allowing outliers or ambiguous examples to dominate optimization, unlike focal loss approaches that monotonically increase weights with difficulty. These harmonized weights are applied to both the numerator and denominator of the Dice loss to account for imbalances caused by variations in $\Delta$.

\begin{equation}
\label{eq:dice_square}
\mathcal{L}_{\text{wDice}} = 1 - \frac{2 \sum_{i=1}^N w_iy_i p_i + \epsilon}{\sum_{i=1}^N w_i(y_i^2 + p_i^2) + \epsilon}
\end{equation}
where \(w_i = w(B_k)\) if \(\Delta_i\) belongs to bin \(B_k\). If the L1 norms of the examples are uniformly distributed, the density \(\mathcal{D}(B_k)\) will be the same for all bins, resulting in equal weights \(w(B_k)\) for all bins. In this case, the weighted DL, (\(L_{\text{wDice}}\)), will reduce to the standard DL.

Finally, we combine \(\mathcal{L}_{\text{wDice}}\) with the Focal Loss. The full expression of L1DFL thus is:

\begin{equation}
\label{eq:l1dfl}
\text{L1DFL} = \mathcal{L}_{\text{wDice}} + \mathcal{L}_{\text{Focal}}
\end{equation}

\begin{figure}[ht]
   \begin{centering}
   \includegraphics[width=13cm]{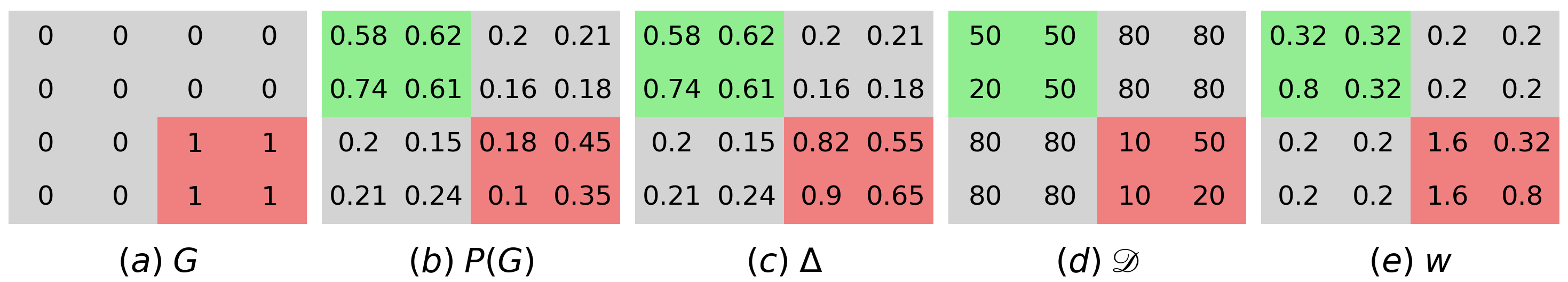}
   \caption{An illustration of the dynamic weighting strategy in L1DFL.(a) Ground truth (G) with lesion areas (1) and background (0). (b) Predicted probability map showing false positives (green) and regions of interest (red). (c) L1 norm values computed as $|\hat{y} - y|$ for each voxel, representing prediction error magnitude. Higher L1 norm values indicate greater classification difficulty. (d) Density $\mathcal{D}$, showing the frequency of each voxel's difficulty level across all voxels. A difficulty level is defined by a range of L1 norm values (bin), and density measures how many voxels fall within that bin. (e) Final weights ($w = N/\mathcal{D}$) assigned through gradient harmonization. Voxels at high-density difficulty levels receive lower weights, while voxels at low-density difficulty levels receive higher weights. This prevents gradient concentration on any single difficulty level, distributing optimization effort more uniformly across the difficulty spectrum.
   \label{fig:scheme} 
    } 
    \end{centering}
\end{figure}

For our implementation, we empirically selected a constant bin width, $\Gamma$, of 0.1 and $\gamma = 2$ for the Focal Loss component of L1DFL. We assess the robustness of the loss function to $\Gamma$ and $\gamma$ hyperparameters, computed over a five-epoch window centered at the epoch with the highest validation Dice. We report the performance in the results section (\ref{ablation}). We illustrate the dynamic weighting strategy of L1DFL in Figure \ref{fig:scheme} using a 4×4 matrix.

\subsection{Empirical Loss Analysis}

To empirically validate the gradient harmonization mechanism of L1DFL and compare it against baseline loss functions, we conducted two complementary analyses on converged models.

\paragraph{Prediction Distribution Analysis:} We evaluated the calibration properties and uncertainty quantification capabilities of each loss function by analyzing the distribution of prediction entropy and confidence scores. For each voxel in the validation set, we computed the prediction entropy as $H = -\sum_{c} p_c \log p_c$ where $p_c$ is the predicted probability for class $c$, and the prediction confidence as $\max_c p_c$. We stratified these distributions by prediction correctness (correct vs. incorrect predictions) to assess whether each loss function produces well-calibrated uncertainty estimates. The \textit{calibration gap}, defined as the difference in mean confidence between correct and incorrect predictions, quantifies how well the model's confidence aligns with its actual accuracy, a larger gap indicates better calibration.

\paragraph{Sample Weighting Analysis:} To directly demonstrate the gradient harmonization effect, we analyzed how each loss function weights samples of varying difficulty. For each test sample, we computed: (1) \textit{sample difficulty} as the mean absolute difference between predictions and targets $d = |\hat{y} - y|$, representing the inherent difficulty of the prediction, and (2) \textit{effective sample weight} as the $L_2$ norm of the loss gradient with respect to the model outputs $w = \|\nabla_{\hat{y}} \mathcal{L}\|_2$, representing how much each sample contributes to the gradient during optimization. We computed the Pearson correlation coefficient $\rho(d, w)$ between difficulty and weight across all samples. A lower correlation indicates that the loss function assigns more balanced weights across the difficulty spectrum, preventing gradient concentration on outliers. We also visualized the relationship between difficulty and weight by binning samples into 50 difficulty levels and computing mean weights per sample for each bin. This analysis reveals whether a loss function over-emphasizes hard samples (high $\rho$) or achieves balanced weighting preventing outlier dominance.

Both analyses were performed on a converged SegResNet model trained with the different loss functions using the validation set, with sliding window inference for computational efficiency.

\subsection{Model Evaluation}
We evaluated the performance of the loss functions on the test set using patient- and lesion-level metrics. Predictions were generated using a sliding-window inference strategy and resampled to the original image coordinates. In addition to overall test-set performance, results were stratified by lesion burden into single-lesion and multiple-lesion scenarios, defined as images containing exactly one lesion and two or more lesions, respectively.

For each scenario and for the overall test set, we performed both patient-level and lesion-level evaluations. At the patient level, we assessed segmentation accuracy using the DSC, detection performance using true positives (TP), false positives (FP), false negatives (FN), and the F1 score, and clinically motivated metrics based on TMTV. For multiple-lesion cases, we additionally evaluated performance with respect to lesion spatial dissemination (\( D_{\text{max}} \)).

Lesion-level evaluation focused on segmentation accuracy (DSC) and performance across different molecular tumor volume (MTV) ranges of individual lesions. Detailed mathematical definitions, matching strategies, thresholding procedures, and clinical metric formulations are provided in \ref{ap:model_evaluation}.

\section{Results}
\label{result}
\subsection{Segmentation performance across different networks}
\label{seg1}

In this section, we present results for the segmentation performance of the three loss functions, DL, DFL, and L1DFL across the different architectures (Table \ref{tab:results}). We report the average and median DSCs at the patient level, with mean values of true positive, false positive, and false negative counts on the test set and the F1 score performance. A one-tailed paired Wilcoxon signed-rank test with Bonferroni correction was performed to evaluate the significance of the performance of L1DFL from those of DL and DFL separately, with corresponding p-values reported.

\begin{table}[ht]
\centering
\resizebox{\textwidth}{!}{
\begin{tabular}{ccccccccc}
\toprule
\multirow{2}{*}{\textbf{Network}} & \multirow{2}{*}{\textbf{Loss Function}} & \multicolumn{2}{c}{\textbf{DSC}} & \multicolumn{3}{c}{\textbf{Mean Rates}} & \multirow{2}{*}{\textbf{F1 scores}} & \multirow{2}{*}{\textbf{P values}} \\
\cmidrule(lr){3-4} \cmidrule(lr){5-7}
 & & \textbf{Mean} & \textbf{Median} & \textbf{TP} & \textbf{FP} & \textbf{FN} & \\
\midrule
\multirow{3}{*}{Attention U-Net} 
 & DL & 0.54 $\pm$ 0.24 & 0.58 [0.44, 0.72] & \textbf{0.81 $\pm$ 0.35} & 2.06 $\pm$ 1.89 & \textbf{0.19 $\pm$ 0.35} & 0.50 $\pm$ 0.28 & $0.004$ \\
 & DFL & 0.51 $\pm$ 0.26 & 0.54 [0.33, 0.73] & 0.77 $\pm$ 0.36 & 2.65 $\pm$ 2.60 & 0.23 $\pm$ 0.36 & 0.44 $\pm$ 0.27 & $<0.001$* \\
 & L1DFL & \textbf{0.58 $\pm$ 0.27} & \textbf{0.66 [0.51, 0.77]} & 0.77 $\pm$ 0.37 & \textbf{0.42 $\pm$ 0.65} & 0.23 $\pm$ 0.37 & \textbf{0.69 $\pm$ 0.34} &  \\
\midrule
\multirow{3}{*}{SegResNet} 
 & DL & 0.52 $\pm$ 0.30 & 0.60 [0.29, 0.76] & 0.73 $\pm$ 0.39 & 0.73 $\pm$ 1.10 & 0.27 $\pm$ 0.39 & 0.62 $\pm$ 0.35 & 0.235 \\
 & DFL & 0.52 $\pm$ 0.25 & 0.59 [0.41, 0.71] & \textbf{0.81 $\pm$ 0.34} & 2.05 $\pm$ 2.00 & \textbf{0.19 $\pm$ 0.34} & 0.49 $\pm$ 0.27 & 0.015 \\
 & L1DFL & \textbf{0.57 $\pm$ 0.29} & \textbf{0.68 [0.47, 0.78]} & 0.76 $\pm$ 0.37 & \textbf{0.52 $\pm$ 0.76} & 0.24 $\pm$ 0.37 & \textbf{0.66 $\pm$ 0.34} & \\
 \midrule
\multirow{3}{*}{U-Net} 
 & DL & 0.49 $\pm$ 0.28 & 0.55 [0.28, 0.75] & 0.69 $\pm$ 0.38 & 1.27 $\pm$ 1.18 & 0.30 $\pm$ 0.39 & 0.49 $\pm$ 0.29 & 0.197\\
 & DFL & 0.49 $\pm$ 0.28 & 0.54 [0.25, 0.70] & \textbf{0.71 $\pm$ 0.39} & 1.66 $\pm$ 1.53 & \textbf{0.28 $\pm$ 0.40} & 0.46 $\pm$ 0.29 & 0.165\\
 & L1DFL & \textbf{0.52 $\pm$ 0.29} & \textbf{0.60 [0.38, 0.76]} & \textbf{0.71 $\pm$ 0.38} & \textbf{0.71 $\pm$ 0.79} & 0.29 $\pm$ 0.39 & \textbf{0.58 $\pm$ 0.33} &  \\
\midrule
\multirow{3}{*}{UNETR} 
 & DL & 0.37 $\pm$ 0.26 & 0.39 [0.11, 0.59] & \textbf{0.64 $\pm$ 0.43} & 7.74 $\pm$ 7.70 & 0.33 $\pm$ 0.42 & 0.21 $\pm$ 0.19 & 0.435\\
 & DFL & \textbf{0.39 $\pm$ 0.28} & 0.46 [0.14, 0.61] & \textbf{0.64 $\pm$ 0.42} & 5.49 $\pm$ 5.77 & \textbf{0.32 $\pm$ 0.40} & 0.26 $\pm$ 0.21 & 0.518\\
 & L1DFL & \textbf{0.39 $\pm$ 0.29} & \textbf{0.48 [0.08, 0.62]} & 0.61 $\pm$ 0.41 & \textbf{2.59 $\pm$ 4.11} & 0.35 $\pm$ 0.42 & \textbf{0.38 $\pm$ 0.29} &  \\
\midrule
\multirow{5}{*}{SAM-Med3D} 
  & Zero-shot & 0.31 $\pm$ 0.24 & 0.26 [0.11, 0.45] & \textbf{0.78 $\pm$ 0.35} & 2.32 $\pm$ 3.39 & 0.33 $\pm$ 0.46 & 0.53 $\pm$ 0.36 & $<0.001$*\\
 & Finetuned (DCE) & \textbf{0.51 $\pm$ 0.25} & \textbf{0.51 [0.29, 0.67]} & 0.74 $\pm$ 0.37 & \textbf{0.45 $\pm$ 1.05} & \textbf{0.29 $\pm$ 0.38} & \textbf{0.71 $\pm$ 0.37} & 0.712\\
 & DL & 0.48 $\pm$ 0.26 & \textbf{0.51 [0.32, 0.67]} & 0.75 $\pm$ 0.37 & 0.90 $\pm$ 3.40 & 0.31 $\pm$ 0.45 & 0.70 $\pm$ 0.38 & 0.290\\
 & DFL & 0.48 $\pm$ 0.26 & \textbf{0.51 [0.30, 0.68]} & 0.73 $\pm$ 0.37 & 0.64 $\pm$ 2.24 & \textbf{0.29 $\pm$ 0.39} & \textbf{0.71 $\pm$ 0.38} & 0.254\\
 & L1DFL & 0.49 $\pm$ 0.26 & 0.50 [0.25, 0.74] & \textbf{0.78 $\pm$ 0.34} & 1.55 $\pm$ 3.56 & 0.33 $\pm$ 0.49 & 0.69 $\pm$ 0.39 & \\
\bottomrule
\end{tabular}
}
\caption{Comparison of the loss functions on the test set based on mean and median patient-level DSC. The mean DSC, TP, FP, FN and F1 scores are presented with standard deviations and the median DSC with interquartile ranges. A detection is considered a true positive if the predicted mask overlaps with the voxel containing the SUVmax value in the ground truth. The P values reflect a one-tailed paired Wilcoxon signed-rank test comparing the median difference in DSC between L1DFL and DL and DFL, with * indicating statistical significance at the $\alpha = 0.003$ level (Bonferroni correction applied).} 
\label{tab:results}
\end{table}

The CNN-based models generally outperformed the transformer-based network and the foundation model, likely due to the limited training data and the inherent inductive bias of CNNs. Across all architectures, median DSC values were consistently higher than the mean DSCs. UNETR exhibited the poorest overall performance, with the DL yielding the lowest DSC and F1 scores, despite achieving competitive mean TP counts. The low DSC and F1 were primarily due to a high FP rate. Similarly, SAM-Med3D showed the lowest DSC in zero-shot testing, although TP rates remained competitive; again, poor performance was linked to elevated FP rates. When fine-tuning SAM-Med3D with different loss functions, including the original Dice Cross-Entropy loss (DCE), median DSCs were similar, with L1DFL lagging by 1 point. However, L1DFL achieved the highest TP counts. Notably, although SAM-Med3D is unimodal and received only PET input, its performance was competitive with multimodal CNN models, likely benefiting from features learned during pretraining.

Within the CNN models, L1DFL consistently achieved the best balance of DSC, FP rate, and F1 score. The combination of L1DFL with SegResNet yielded the highest median DSC $(0.68)$, whereas the best mean TP counts were obtained by DL and DFL on SegResNet and Attention U-Net, respectively. However, these high sensitivity values came at the cost of increased false positives, reducing F1 scores. For example, DL on SegResNet had an FP rate of $0.73\pm1.10$, whereas on Attention U-Net it had $2.06\pm1.89$, roughly three times higher. Overall, Attention U-Net and SegResNet outperformed U-Net. DL consistently achieved higher median DSC, FP, and F1 scores than DFL across all CNNs, although DFL had higher TP counts on SegResNet and U-Net. 

For the transformer-based model, DFL slightly outperformed DL in FP and F1 scores. L1DFL, however, achieved the best balance of TP detection and minimal FPs on both CNNs and UNETR, and showed higher median DSC in the third quartile range for all models. For instance, on UNETR where DL and DFL had mean FP rates above 5.0, L1DFL minimized false detections while maintaining lesion sensitivity, yielding the highest mean and median DSCs for this network.

Although a one-tailed Wilcoxon signed-rank test comparing loss functions across networks did not provide sufficient evidence to reject the null hypothesis, statistically significant differences were observed on Attention U-Net when comparing L1DFL to DFL. The training and validation dynamics of all networks and loss functions are provided in \ref{ap:train_curve}, showing consistent convergence across models.

 \begin{figure}[tb!]
   \begin{centering}
   \includegraphics[width=\linewidth, keepaspectratio]{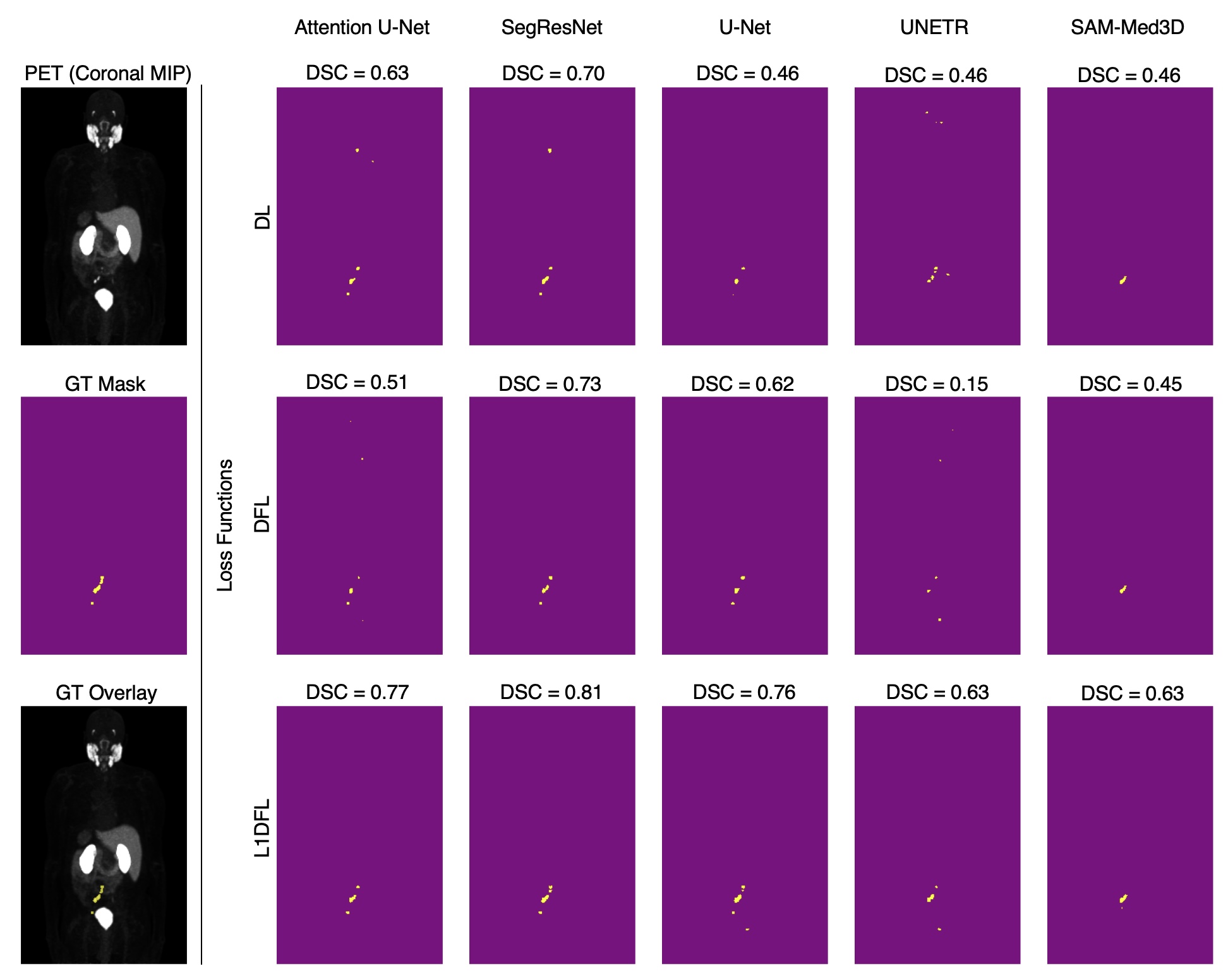}
   \caption{Segmentation results on a test PET image. The first column shows the coronal maximum intensity projection (MIP), ground truth (GT) mask, and GT overlaid on the PET image. Columns correspond to different networks, and rows correspond to different loss functions.
   \label{fig:segment} 
    } 
    \end{centering}
\end{figure}

Figure \ref{fig:segment} illustrates the segmentation performance of the different loss functions on a test PET image. In this example, SegResNet achieved the most accurate lesion segmentation with the highest DSC among all models. DL generally outperformed DFL, except for SegResNet. Across all networks, L1DFL provided the best performance, including on the transformer-based models where DL and DFL struggled.

\subsection{Performance in single and multiple lesion scenarios}
\label{scenarios}
\begin{figure}[tb!]
   \begin{centering}
   \includegraphics[width=\linewidth, keepaspectratio]{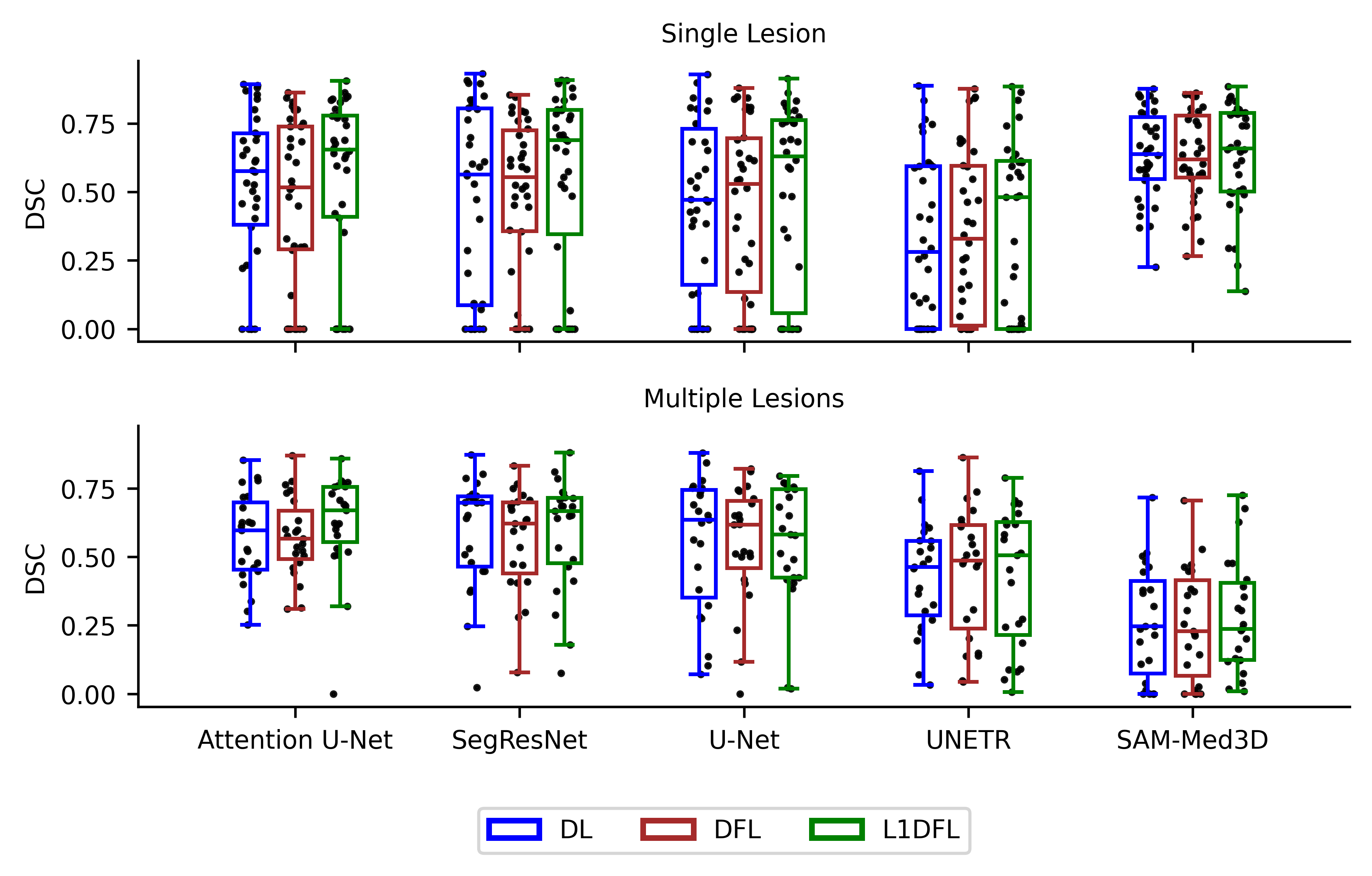}
   \caption{Box plots overlaid with swarm plots illustrating the distribution of Dice Similarity Coefficient (DSC) performance for the three loss functions across two lesion scenarios: Multiple lesions (lesion count $> 2$) and Single lesion (lesion count $= 1$). Each box plot's top and bottom edges represent the interquartile range (IQR), while the horizontal lines within the boxes indicate the median DSC values. Whisker lengths are set to 1.5 times the IQR. The number of cases in the single lesion scenario was N = 34 and for the multiple lesion scenario, the number of cases was N = 23.
   \label{fig:sing_multi_dsc} 
    } 
    \end{centering}
\end{figure}

We evaluated the loss functions in two lesion scenarios: single lesion and multiple lesions in an image. Figure \ref{fig:sing_multi_dsc} shows that DSC values were more widely spread in the single-lesion scenario across all networks. The largest interquartile range (IQR) differences between scenarios were observed for UNETR across all loss functions and SegResNet with DL, whereas Attention U-Net and SAM-Med3D maintained tighter IQRs. Additionally, DSC values beyond the 75th percentile were generally higher in the single-lesion scenario than the multiple lesion scenario, with more cases exceeding $0.75$.

For the single lesion scenario, L1DFL performed consistently better than the other loss functions. Specifically, on the CNN models, L1DFL improved median DSC by at least $19\%$ over DL and DFL, while for UNETR and SAM-Med3D, improvements were at least $45\%$ and $3.3\%$, respectively. In contrast, DL and DFL showed variable performance: DL outperformed DFL on Attention U-Net, SegResNet, and SAM-Med3D, while DFL achieved higher median DSC on U-Net and UNETR. L1DFL’s best performance was on SegResNet with a median DSC of $0.69$, whereas DL and DFL performed best on SAM-Med3D with a median DSC of $0.64$ and $0.62$, respectively.

In the multiple-lesion scenario, even though DL generally achieved the best overall performance across the networks, L1DFL had the highest median DSCs on UNETR and Attention U-Net. Conversely, SAM-Med3D struggled to achieve accurate lesion segmentation in this scenario, and although its overall median DSC (Table \ref{tab:results}) was higher than UNETR, UNETR performed better at segmenting multiple lesions. For SAM-Med3D, all loss functions had median DSCs in the range $[0.22, 0.25]$, with tighter IQRs suggesting generally poor performance on cases with multiple lesions in the image.

\subsection{Performance based on lesion molecular volume}
\label{mtv}

\begin{figure}[tb!]
   \begin{centering}
    \includegraphics[width=\linewidth, keepaspectratio]{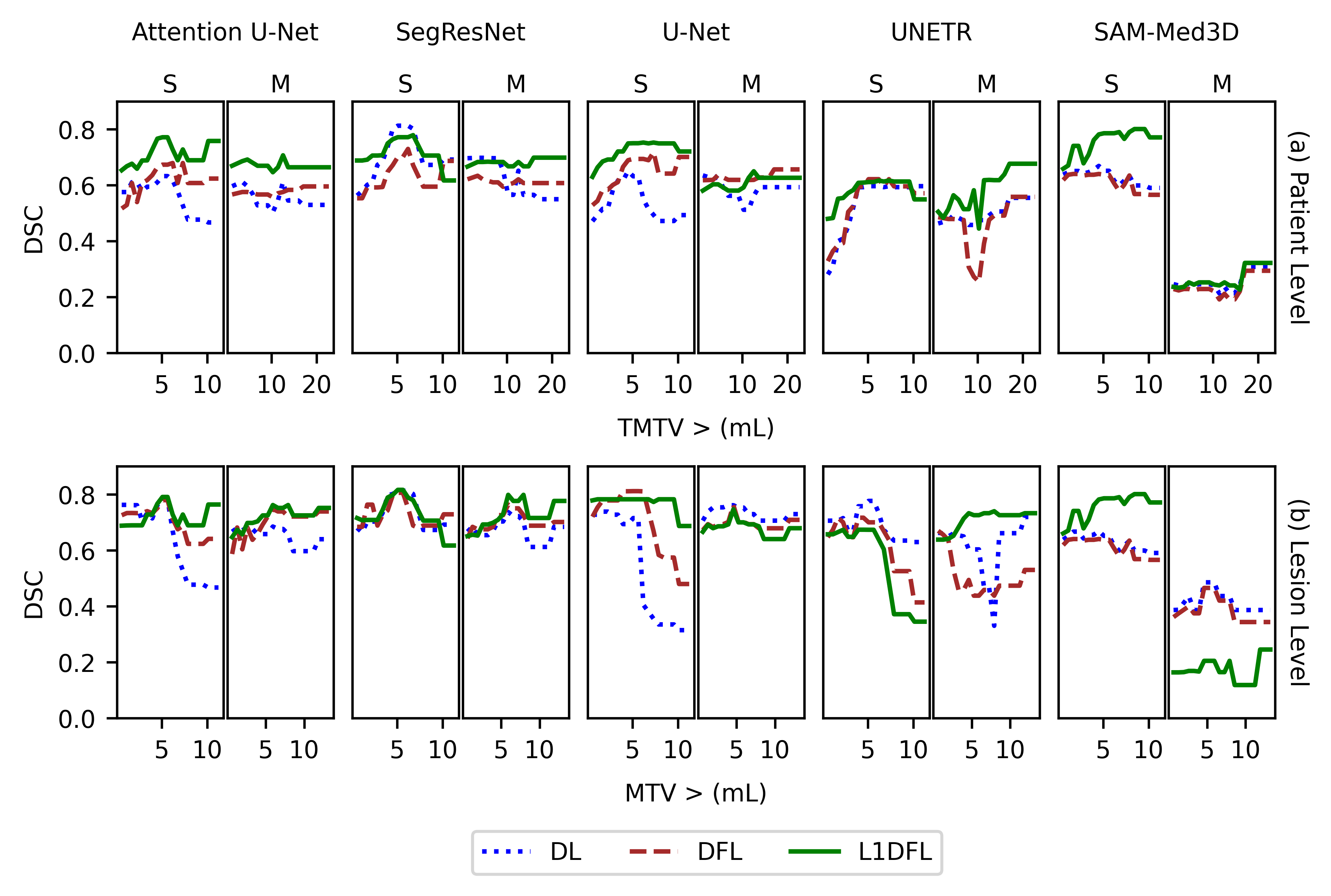}
   \caption{Patient- and lesion-level median Dice Similarity Coefficient (DSC) performance of the loss functions across varying ground truth tumor volumes. Each column corresponds to a network, with subcolumns showing single-lesion (S) and multiple-lesion (M) scenarios. The top row presents patient-level results versus total molecular tumor volume (TMTV), while the bottom row shows lesion-level results versus molecular tumor volume (MTV). Thresholds $(t)$ span the interquartile range up to the 85th percentile, and median DSC is computed for all lesions $(l)$ with volume $(v_l > t)$. Line colors indicate different loss functions.
   \label{fig:tmtv_trend} 
    } 
    \end{centering}
\end{figure}

We also assessed how the loss functions performed across varying tumor volumes in both single- and multiple-lesion scenarios (Figure \ref{fig:tmtv_trend}). DL yielded higher DSCs for smaller lesions, whereas DFL was slightly more robust to lesion size variations, especially on CNN-based models. Nevertheless, performance for both loss functions generally declined with increasing tumor volume. For instance, in single-lesion scenarios, DL and DFL achieved higher DSCs at lower volumes, but their performance decreased more sharply at larger volumes, occasionally resulting in relatively better outcomes for multiple-lesion cases. Besides, the segmentation accuracy for both loss functions was higher at the lesion-wise assessment level than at the patient level for smaller tumors, likely due to the greater impact of false positives.

L1DFL, in contrast, demonstrated a consistent advantage over DL and DFL, particularly at the patient level, where it exhibited an upward DSC trend with increasing TMTV across all networks and remained stable across different lesion scenarios. Moreover, on the CNN-based models, L1DFL consistently maintained DSC values between $0.6$ and $0.8$ across both patient- and lesion-level assessments, whereas DL and DFL either started within this range and declined as volume increased or began lower and gradually rose into it.

From a network perspective, CNNs generally achieved higher segmentation performance than UNETR and SAM-Med3D, with Attention U-Net and SegResNet performing best among the CNNs, and U-Net performing less favorably. DL paired well with SegResNet, while DFL produced similar results on SegResNet and Attention U-Net. Both losses, however, showed reduced performance on U-Net, particularly for larger lesions; for example, in the single-lesion scenario beyond an MTV threshold of $5$ mL, DL’s median DSC decreased from $0.73$ to $0.32$, and DFL from $0.81$ to $0.48$. L1DFL, on the other hand, remained relatively stable, with only a $14\%$ decrease. On Attention U-Net, DL decreased by $66\%$ at high MTV values and DFL by $21\%$, while L1DFL exhibited only a slight dip with a mild upward trend overall.

UNETR and SAM-Med3D exhibited more variable performance across assessment levels. On UNETR, DL and DFL declined in performance with increasing tumor volumes, whereas L1DFL remained robust, showing a downward trend only in lesion-wise analysis for single-lesion cases. SAM-Med3D performed better for single-lesion cases than UNETR, although DL and DFL still exhibited decreasing performance. In contrast, for multiple-lesion cases, SAM-Med3D showed a sharp drop in segmentation accuracy. Overall, finetuning SAM-Med3D with all three loss functions did not improve performance on the multiple lesions cases, with L1DFL trailing DL and DFL by at least $40\%$.

\subsection{Performance based on lesion dissemination}
\label{Dmax}
\begin{figure}[tb!]
   \centering
   \includegraphics[width=\linewidth, keepaspectratio]{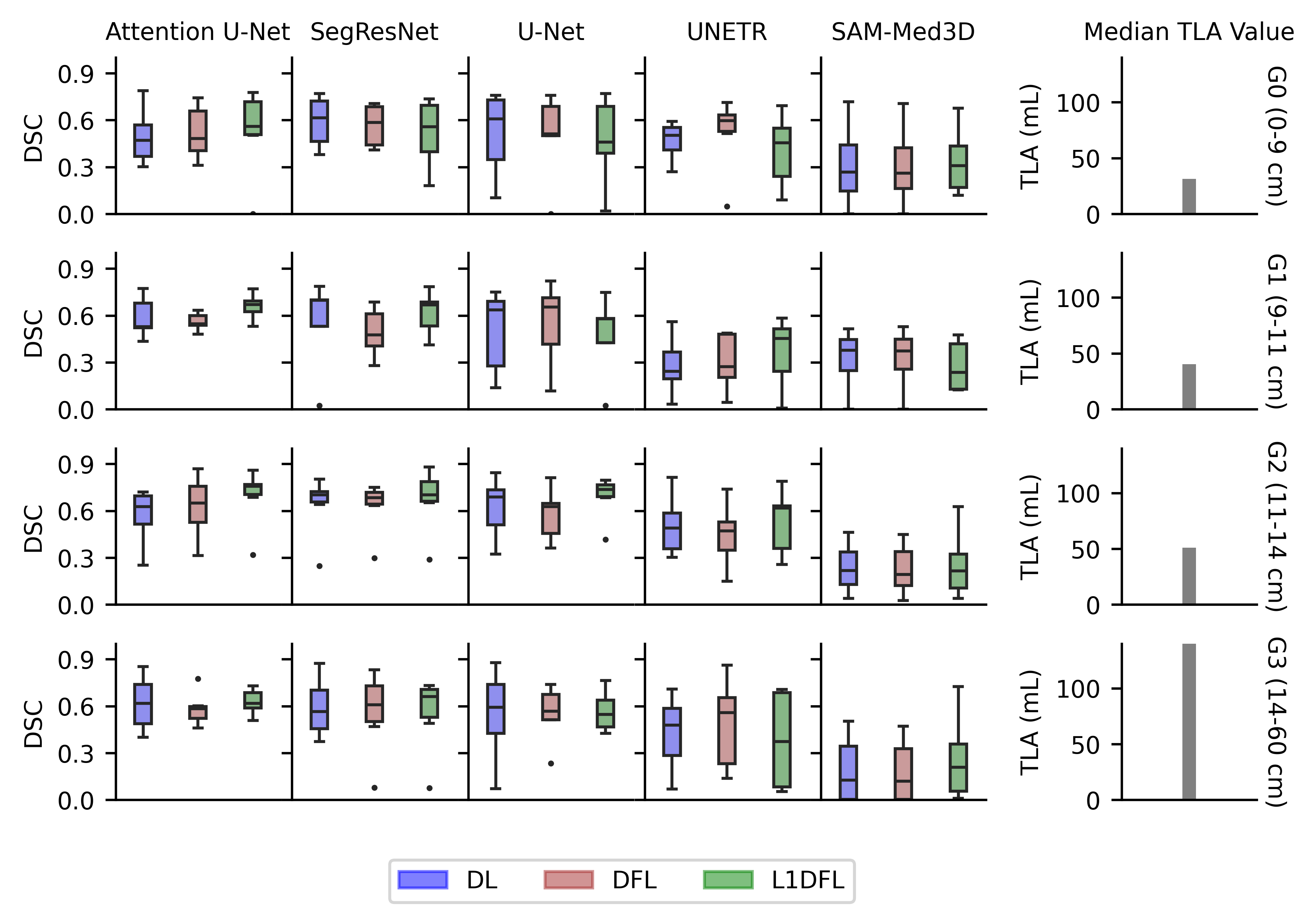}
   \caption{Plots illustrating the Dice Similarity Coefficients (DSC) across four interquartile ranges of \( D_{\text{max}} \). The groups G0 (0-9 cm), G1 (9-11cm), G2 (11-14cm), and G3 (14-60cm) correspond to the interquartile ranges of $0-25\%$, $25-50\%$, $50-75\%$, and $75-100\%$, respectively. Each column show performance on a specific network. In the box plots, horizontal lines indicate the median, and whisker lengths are set to 1.5 times the interquartile range (IQR), with outliers represented as black dots. Accompanying bar plots in the last column display the median total lesion activity (TLA) values for cases within each \( D_{\text{max}} \) group. The different loss functions are color-coded for clarity. The total number of cases in each group was N = 6.
   \label{fig:dmax}
    } 
\end{figure}

To quantify the spatial distribution of lesions, we calculated \( D_{\text{max}} \), the maximum Euclidean distance between any two lesions in the image. For the performance assessment of the loss functions based on the \( D_{\text{max}} \), we categorized the distances into the following groups: 0-9 cm (G0), 9-11 cm (G1), 11-14 cm (G2), and 14-60 cm (G3), representing the first, second, third and fourth quartile ranges, respectively. Patient-level analysis of DSC showed consistent improvement across these ranges (Figure \ref{fig:dmax}) with differences in segmentation performance across architectures. CNN-based models generally performed better on widely spread lesions (G2 and G3) than on closer distances, although performance declined from G2 to G3, indicating increasing challenges as tumors extend further. While SAM-Med3D and UNETR performed worse overall than the CNN-based models, they showed relatively stronger performance at lower \( D_{\text{max}} \) ranges, with accuracy decreasing as lesions spread further, even in the presence of higher tumor activity indicated by the TLA plots.

Moreover, performance generally varied with the choice of loss function for each network. At lower distances (G0 and G1), DL with SegResNet achieved the highest median DSCs, while L1DFL paired with Attention U-Net offered greater consistency. At higher distances (G2 and G3), L1DFL with Attention U-Net or SegResNet provided the strongest and most consistent results among the CNN models, with the widest IQR being $0.17$. Additionally, U-Net with L1DFL was competitive at G2, reaching a median DSC of $0.74$ and outperforming DL and DFL on all three CNN models by at least $5\%$. DL, on the other hand, performed best in the first and fourth quartiles on U-Net, while DFL yielded improved performance in the second and third quartiles. Overall, Attention U-Net with L1DFL provided the most consistent performance across \( D_{\text{max}} \) ranges while SegResNet was most robust at larger lesion spreads. U-Net performed best at lower distances.

On SAM-Med3D, all loss functions showed reduced performance, with median DSCs ranging from $0.20 \text{ – } 0.40$ at lower lesion spreads (G0 and G1) and dropping to $0.10\text{ – }0.25$ at higher spreads (G2 and G3). Despite the generally poor results, DL and DFL achieved a $40\%$ improvement over UNETR in the second quartile, although their performance remained below CNN-based models. In contrast, L1DFL performed better when paired with UNETR than with SAM-Med3D, albeit with less consistency as lesion spread increased, as indicated by the wider IQR. DFL showed the highest performance on UNETR in the first and last quartiles, while L1DFL performed best at G1 and G2.

\subsection{Loss Function Properties}

\subsubsection{Prediction Distribution Analysis}

From Table \ref{tab:loss_analysis}, L1DFL demonstrates better calibration than the baseline methods. Its calibration gap (0.2859) is substantially larger than that of DFL (0.0263) and DL (0.0129), indicating stronger separation between correct and incorrect predictions. When errors occur, L1DFL appropriately expresses uncertainty, with a mean entropy of 0.4762 for incorrect predictions versus 0.0212 for correct ones. In contrast, DFL and DL remain highly overconfident even on incorrect predictions, with confidence scores of 0.9737 and 0.9871, respectively.

\begin{table}[ht]
\centering
\caption{Quantitative comparison of loss function properties}
\label{tab:loss_analysis}
\resizebox{\textwidth}{!}{%
\begin{tabular}{lcccccc}
\toprule
& \multicolumn{3}{c}{\textbf{Prediction Distribution}} & \multicolumn{3}{c}{\textbf{Sample Weighting}} \\
\cmidrule(lr){2-4} \cmidrule(lr){5-7}
\textbf{Loss Function} & 
\makecell{Entropy\\(correct)} & 
\makecell{Entropy\\(incorrect)} & 
\makecell{Calibration\\Gap} & 
\makecell{Confidence\\(correct)} & 
\makecell{Confidence\\(incorrect)} & 
\makecell{Difficulty-Weight\\Correlation} \\
\midrule
L1DFL & 0.0212 & 0.4762 & 0.2859 & 0.9982 & 0.7123 & 0.3376 \\
DFL & 0.0000 & 0.0614 & 0.0263 & 1.0000 & 0.9737 & 0.9587 \\
DL & 0.0000 & 0.0304 & 0.0129 & 1.0000 & 0.9871 & 0.0961 \\
\bottomrule
\end{tabular}%
}
\end{table}

\begin{figure}[tb!]
   \centering
   \includegraphics[width=\linewidth, keepaspectratio]{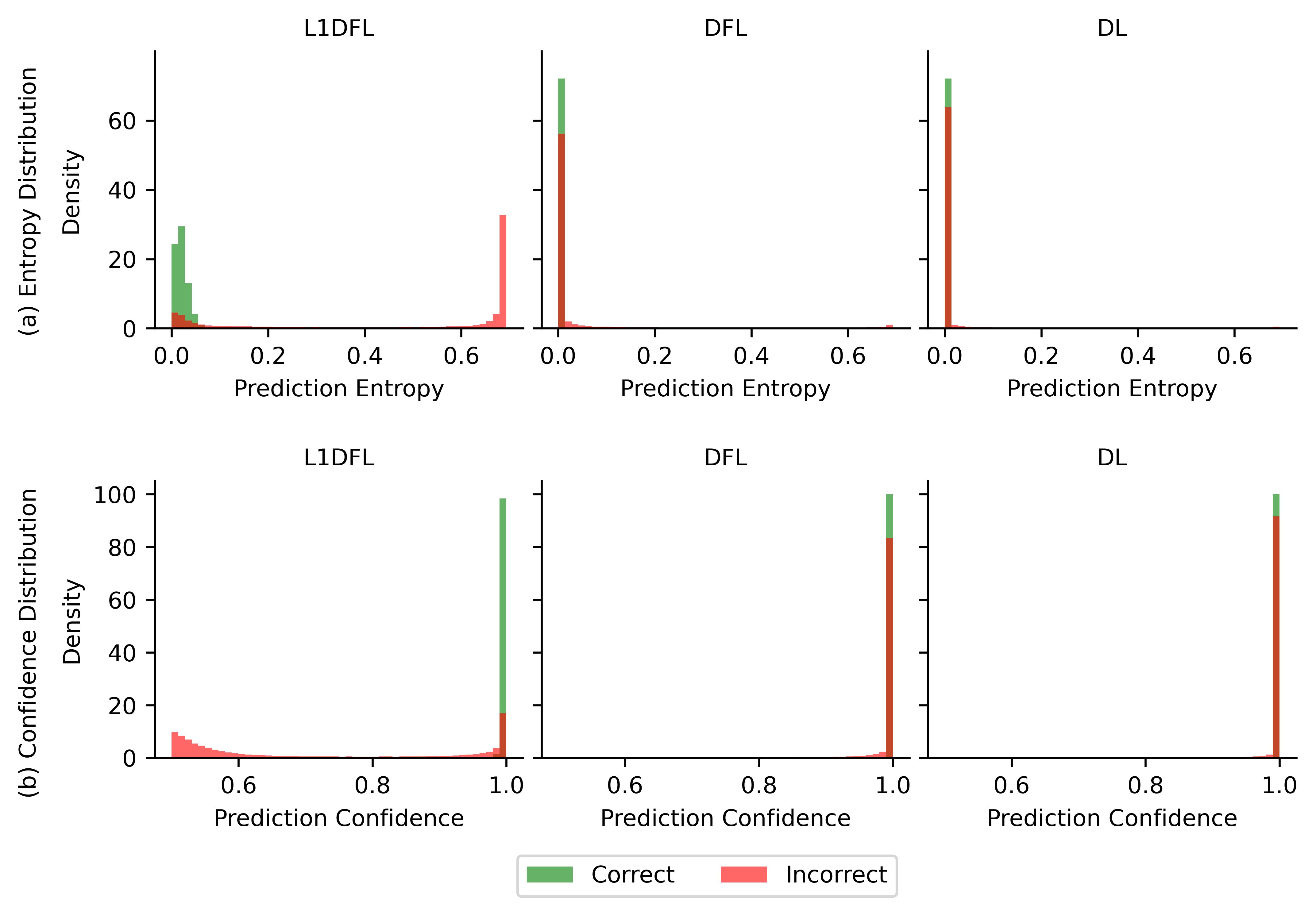}
   \caption{Entropy (top) and confidence (bottom) distributions for correct and incorrect predictions. The model trained with L1DFL exhibits clear separation between confident correct predictions and uncertain incorrect predictions, whereas DFL and DL remain overconfident regardless of prediction correctness.
   \label{fig:prediction_dist}
    } 
\end{figure}

These trends are reflected in the entropy distributions (Figure \ref{fig:prediction_dist}, top row). L1DFL exhibits a clear bimodal structure, with correct predictions concentrated at near-zero entropy and incorrect predictions at higher entropy values, indicating the model appropriately recognizes ambiguous cases. Conversely, DFL and DL collapse to near-zero entropy for both correct and incorrect predictions, demonstrating poor uncertainty discrimination.

Similarly, the confidence distributions (Figure \ref{fig:prediction_dist}, bottom row) show that L1DFL assigns lower confidence to incorrect predictions while maintaining high confidence for correct ones. In contrast, DFL and DL produce near-saturated confidence values for all predictions, offering little discriminative value for uncertainty estimation.

\begin{figure}[tb!]
   \begin{centering}
   \includegraphics[width=11cm, keepaspectratio]{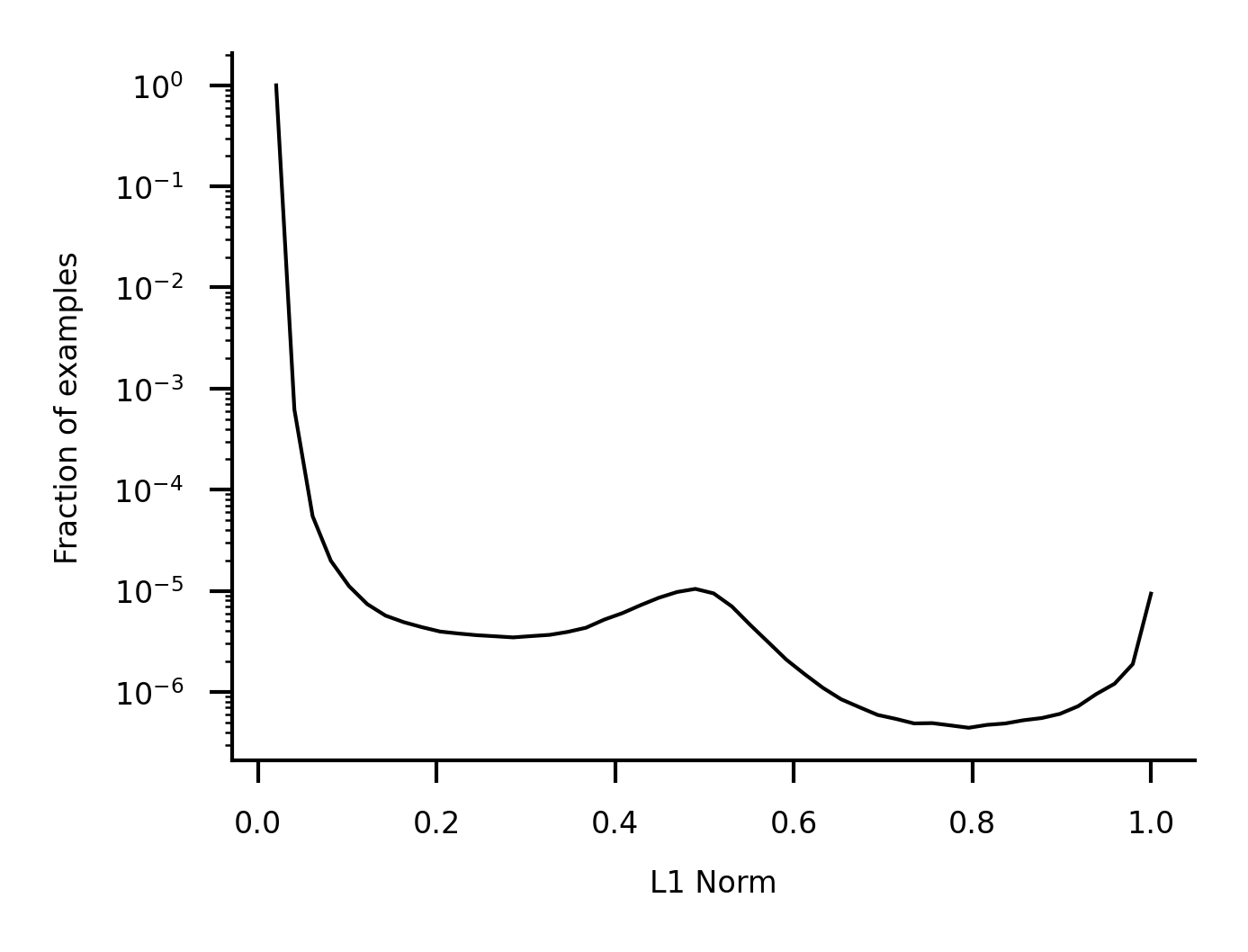}
   \caption{L1-norm distribution of a converged model trained with L1DFL (log-scaled y-axis).
   \label{fig:fractions} 
    } 
    \end{centering}
\end{figure}

Figure \ref{fig:fractions} illustrates the L1-norm distribution of a converged SegResNet trained with L1DFL. As expected, the fraction of easy examples is large and could dominate the overall gradient. However, a non-negligible number of very difficult examples remain misclassified. As noted by Li et al. \cite{li2019gradient}, such samples can be treated as outliers, and forcing the model to fit them may degrade performance on the majority of data due to their inconsistent gradient directions. In addition, a substantial fraction of samples lies near the decision boundary, reflecting intrinsic ambiguity or class overlap. The prominence of this ambiguous region indicates that the model captures uncertainty rather than enforcing overconfident predictions, consistent with the bimodal entropy and confidence distributions shown in Figure \ref{fig:prediction_dist}.

\subsubsection{Sample Weighting Analysis}

\begin{figure}[tb!]
   \centering
   \includegraphics[width=\linewidth, keepaspectratio]{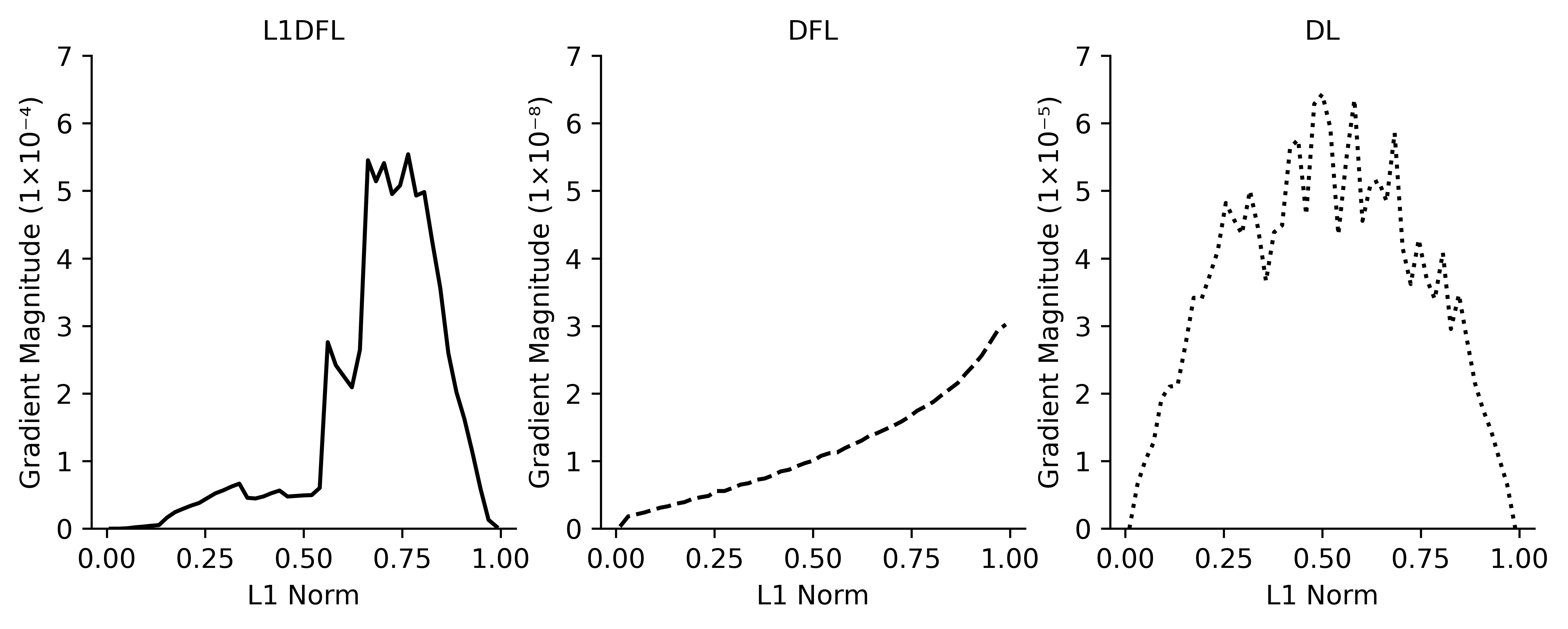}
   \caption{Comparison of sample weighting across difficulty levels for the different loss functions. L1DFL emphasizes moderately difficult samples, whereas DFL heavily weights the hardest examples and DL shows inconsistent weighting. 
   \label{fig:sample_weights}
    } 
\end{figure}

Figure \ref{fig:sample_weights} provides direct evidence of L1DFL's gradient harmonization mechanism through analysis of effective sample weighting across difficulty levels. The difficulty-weight correlation for L1DFL $(0.3376)$ is substantially lower than DFL $(0.9587)$ indicating that L1DFL assigns more uniform weights across the difficulty spectrum rather than exhibiting strong monotonic dependence on sample difficulty. DL on the other hand has the lowest correlation of $0.0961$ indicating a difficulty agnostic behavior in weighting.

Figure \ref{fig:sample_weights} shows that L1DFL assigns relatively balanced weights across difficulty levels, with a controlled emphasis on moderately difficult samples in the $0.6$–$0.8$ range, consistent with its intended gradient harmonization behavior. In contrast, DFL applies monotonically increasing weights, with gradient magnitudes growing rapidly as difficulty approaches 1.0, causing optimization to be dominated by the hardest samples, including outliers and ambiguous cases. DL exhibits more irregular weighting patterns, lacking a clear and consistent structure across difficulty levels.

\subsection{Ablation Study and Hyperparameter Robustness Test}
\label{ablation}

To assess the contribution of each component of the proposed L1DFL loss, we conducted an ablation study on two representative architectures, SegResNet and UNETR. Starting from a Dice+Focal baseline, we progressively modified the voxel-wise weighting strategy by (i) replacing the histogram-based weighting with raw gradient magnitudes, (ii) using squared gradients, and (iii) applying the full proposed formulation. All experiments were performed using the same training protocol and hyperparameters.

\begin{table}[H]
\centering
\caption{Ablation study evaluating the impact of individual components of the proposed L1DFL loss on validation Dice score (mean $\pm$ std).}
\begin{tabular}{lcc}
\hline
Loss Variant & SegResNet & UNETR \\
\hline
Dice + Focal (baseline) & 0.50 $\pm$ 0.02 & 0.33 $\pm$ 0.02 \\
Only gradients ($w=\Delta$) & 0.50 $\pm$ 0.02 & 0.35 $\pm$ 0.02 \\
Gradients squared ($w=\Delta^2$) & 0.47 $\pm$ 0.03 & 0.33 $\pm$ 0.02 \\
L1DFL (proposed) & \textbf{0.53 $\pm$ 0.02} & \textbf{0.36 $\pm$ 0.02} \\
\hline
\label{tab:ablation}
\end{tabular}
\end{table}

As shown in Table \ref{tab:ablation}, using raw gradients provides a small improvement over the baseline for UNETR but has little effect for SegResNet. Squaring the gradients ($w=\Delta^2$) reduces performance for both architectures, highlighting the importance of controlled weighting. The full L1DFL formulation achieves the highest Dice scores for both networks, demonstrating that each component, including the histogram-based gradient weighting and the L1 normalization, contributes to improved segmentation performance.

We conducted a validation-based hyperparameter sensitivity analysis. Table \ref{tab:hyperparam} reports the mean and standard deviation of the validation Dice across a range of $\Gamma$ and $\gamma$ values. Performance varies slightly across configurations, ranging from $0.51$ to $0.54$, with overlapping standard deviations, indicating that L1DFL is robust to these hyperparameter settings. We present the results using the SegResNet model; however, comparable ranges of Dice score differences were also observed with SAM-Med3D and UNETR.

\begin{table}[ht]
\centering
\caption{Validation-set Dice scores for different $\gamma$ (Focal Loss) and bin width $\Gamma$ values. Values are reported as mean $\pm$ standard deviation over a 5-epoch window centered at the best validation epoch.}

\label{tab:hyperparam}
\begin{tabular}{c|c|c}
\toprule
$\gamma$ & $\Gamma$ & Dice \\
 \midrule
1 & 0.05 & 0.53 $\pm$ 0.03 \\
1 & 0.10 & 0.51 $\pm$ 0.02 \\
1 & 0.20 & 0.51 $\pm$ 0.04 \\
 \midrule
2 & 0.05 & 0.53 $\pm$ 0.04 \\
2 & 0.10 & 0.53 $\pm$ 0.02 \\
2 & 0.20 & 0.54 $\pm$ 0.03 \\
 \midrule
5 & 0.05 & 0.53 $\pm$ 0.03 \\
5 & 0.10 & 0.53 $\pm$ 0.02 \\
5 & 0.20 & 0.54 $\pm$ 0.01 \\
\bottomrule
\end{tabular}
\end{table}

\section{Discussion}
\label{discuss}

In this work, we proposed and evaluated a novel loss function L1DFL, in segmenting metastatic prostate cancer lesions based on different lesion scenarios characterized by the number of lesions in an image. Specifically, we defined two scenarios: single lesion scenario which include whole body PET/CT images with only a single lesion present in the ground truth mask and multiple lesion scenario being images with more than one lesion. We compared the performance of our proposed loss function with the DL and DFL. Moreover, we report how the size of a lesion and its spatial distribution influence the segmentation result. The dataset used in this study included tumors which had resurfaced after treatment with relatively low molecular volumes and standardized uptake values, thus introducing unique challenges for correct segmentations.

The CNN-based models achieved better overall segmentation performance, with higher true positive detections and median DSCs compared to UNETR and SAM-Med3D. This outcome is expected, as the limited training data favored CNNs, whose inductive bias enables them to learn relevant features more effectively from smaller datasets than transformer-based models \cite{lu2022bridging, lee2022towards}. Nevertheless, SAM-Med3D, as a pretrained medical foundation model, demonstrated competitive results when fine-tuned on the limited dataset. Our findings further showed that the performance of the loss functions is not uniform across different architectures. For instance, the results show that the DL generated more false positives on Attention U-Net compared to SegResNet, despite showing higher sensitivity on the former. Similarly, the DFL exhibited lower precision on the Attention U-Net architecture due to increased false positives. Both loss functions also had lower F1 scores on U-Net suggesting a lower balance of true positives and false positives. By evaluating performance at both the patient and lesion-levels, the effect of false positive rates was highlighted. There was a more than $10\%$ decrease in the median DSC values at the lesion level for both loss functions when assessments were made at the patient level, especially for lower tumor volume thresholds. L1DFL, in contrast, was more robust to these false detections. 

Additionally, by assessing performance based on the different lesion scenarios, the influence of various lesion characteristics such as volume, extent of spread, number of lesions, on accuracy could be evaluated. For instance, considering the single lesion scenario, the performance of the DL is shown to decline as the threshold volume of tumors increased beyond $5 \text{ml}$. This could suggest that DL is highly sensitive to the voxels of high uptake values \cite{XU2023106882}. This observation aligns with claims in \cite{LIU2024103015} where it is indicated that the DL is inherently biased toward smaller regions and may generally perform poorly on larger lesions with more variable SUV distribution across the volume. 

The DL is based on overlap measurement and it inherently balances precision with recall, making it more robust in scenarios with inter-class class imbalances as small objects tend to acquire greater contribution to the overall loss \cite{sudre2017generalised, LIU2024103015}. However, it does not distinguish between intra-class imbalances leading to a balanced focus on both easy and difficult examples \cite{Kofler2023BlobLoss, 9338261}. These two imbalance scenarios tend to dominate among recurrent prostate cancer images, consequently, at lower tumor volumes, the DL had higher DSC values than the DFL, yet struggles to maintain a consistently higher performance on scenarios of multiple lesions with larger volumes. This trend, however, is not limited to the DL, as slight performance declines were observed in the multiple-lesion setting, particularly at the patient assessment level for all losses, indicating added complexity as lesion count increases. In fact, with more lesions, different anatomical sites may be involved, and the effect of other organs with high uptake values could be pronounced \cite{XU2023106882}. Besides, the evaluation based on lesion spread, \(D_{\text{max}}\), highlights the decrease in the models' performance as distribution across the body becomes more extensive.

Overall, L1DFL produced the most robust segmentations across evaluation scenarios. Its superior performance stems from a gradient harmonization mechanism that mitigates two central challenges in imbalanced medical image segmentation: dominance of easy background voxels and instability caused by outlier samples. By estimating mini-batch statistics of sample difficulty (via the L1 norm) and difficulty density, L1DFL redistributes gradient contributions such that frequent difficulty levels are down-weighted while underrepresented ones are emphasized. This prevents training from being dominated by either abundant background voxels or extreme cases \cite{li2019gradient}.

Our analysis showed that extreme-difficulty samples (L1 norms of $0.9$ – $1.0$) persist after convergence and likely correspond to outliers, ambiguous annotations, or imaging artifacts. These samples occur more frequently than moderately difficult ones and may introduce inconsistent gradient directions, destabilizing optimization \cite{li2019gradient}. By down-weighting them according to their prevalence, L1DFL reduces sensitivity to noisy labels and spurious correlations.

Empirically, L1DFL also yields a bimodal prediction-entropy distribution, clearly separating confident correct predictions from appropriately uncertain ones. This behavior is particularly valuable in clinical settings where reliable uncertainty estimation is essential \cite{begoli2019need}. In contrast, although the DL weights samples equally irrespective of the difficulty level \cite{9338261} and the Focal Loss encourages the model to focus more on hard samples \cite{lin2017focal}, the DFL tends to lose the DL component \cite{YEUNG2022102026}, thus, assigns monotonically increasing weights with difficulty, overemphasizing extreme cases regardless of prevalence, leading to overconfident errors.

Moreover, the proposed loss function introduces additional hyperparameters ($\gamma$ for focal loss, $\Gamma$ for bin width) that require tuning. While we show robustness to these parameters in this study, optimal values may vary across datasets and applications. Also, L1DFL incurs higher computational cost than standard losses due to computing L1 norms, binning samples, calculating density, and applying density-based weights. While this overhead remains practical for medical image segmentation tasks in our experiments, it may be non-negligible for extremely large 3D volumes. The benefits of gradient harmonization are most pronounced in highly imbalanced scenarios with diverse difficulty distributions. For a relatively balanced distribution, the L1DFL reduces to the Dice Focal loss with no added advantage to the additional complexity. Additional limitations of this study include the dataset being restricted to patients with up to five PSMA-avid lesions, representing a consecutive oligometastatic BCR population; thus, the performance of the loss functions in more extensive disease remains untested. Although improvements were observed across evaluation metrics, most did not reach statistical significance. Inter-observer variability in the ground truth annotations was not assessed, so segmentation errors may have influenced model performance. Finally, the generalizability of L1DFL to other medical imaging modalities remains uncertain, as the analysis was limited to PET/CT scans.

\section{Conclusion}
\label{conc}

We evaluated the performance of a novel loss function, \textit{L1DFL}, in the detection and segmentation of metastatic prostate cancer lesions in PSMA PET/CT scans across single and multiple lesion scenarios, characterized by lesion count. L1DFL employs a gradient harmonization scheme that leverages sample difficulty, quantified by L1 norms, and the density of difficulty levels to modulate the Dice Loss, which is then combined with Focal Loss to emphasize challenging voxel classifications. We compared L1DFL against DL and DFL across three 3D CNN architectures (Attention U-Net, SegResNet, and U-Net), a transformer-based model (UNETR), and a medical foundation model (SAM-Med3D). While DL generally showed higher sensitivity to small lesions, both DL and DFL exhibited variable performance on larger volumes and more extensive lesion spread, with DFL in particular producing higher false positive rates. In contrast, L1DFL led to balanced segmentation performance across lesion scenarios and architectures, achieving higher F1 scores and Dice Similarity Coefficients. Future work will be directed at exploring the generalizability of L1DFL to other medical imaging datasets and tasks.

\section{Acknowledgments}
This work was supported by the Canadian Institutes of Health Research (CIHR) Project Grants PJT-162216 and PJT-173231. We also acknowledge computational resources and services provided by Microsoft AI for Health.

\section{Ethics Statement}
The datasets used in this work included 18F-DCFPyL PET/CT scans of patients diagnosed with biochemical recurrence of prostate cancer as part of an ongoing clinical trial (ClinicalTrials.gov NCT02899312) with informed written consent obtained from the participants. The ethics approval was granted by the UBC BC Cancer Research Ethics Board (REB) (REB Number: H16-01551).

\section{Declaration of Competing Interest}
The authors declare that they have no known competing personal or financial interests that could have influenced this work.

\section{Declaration of generative AI and AI-assisted technologies in the writing process}
During the preparation of this work the author used OpenAI GPT o1 in order to refine statements. After using this tool, the author reviewed and edited the content as needed and takes full responsibility for the content of the publication.

\section{Data and Code Availability}
The dataset used in this study is proprietary and cannot be publicly released due to data privacy restrictions. However, the implementation code used to develop and evaluate the proposed method is publicly available at: https://github.com/ObedDzik/pca\_segment.git.
\bibliography{reference}

@INPROCEEDINGS{9338261,
  author={Zhao, Rongjian and Qian, Buyue and Zhang, Xianli and Li, Yang and Wei, Rong and Liu, Yang and Pan, Yinggang},
  booktitle={2020 IEEE International Conference on Data Mining (ICDM)}, 
  title={Rethinking Dice Loss for Medical Image Segmentation}, 
  year={2020},
  volume={},
  number={},
  pages={851-860},
  doi={10.1109/ICDM50108.2020.00094}}

@article{LIU2024103015,
title = {Do we really need dice? The hidden region-size biases of segmentation losses},
journal = {Medical Image Analysis},
volume = {91},
pages = {103015},
year = {2024},
issn = {1361-8415},
doi = {https://doi.org/10.1016/j.media.2023.103015},
url = {https://www.sciencedirect.com/science/article/pii/S136184152300275X},
author = {Bingyuan Liu and Jose Dolz and Adrian Galdran and Riadh Kobbi and Ismail {Ben Ayed}}}

@InProceedings{10.1007/978-3-319-46976-8_19,
author="Drozdzal, Michal
and Vorontsov, Eugene
and Chartrand, Gabriel
and Kadoury, Samuel
and Pal, Chris",
editor="Carneiro, Gustavo
and Mateus, Diana
and Peter, Lo{\"i}c
and Bradley, Andrew
and Tavares, Jo{\~a}o Manuel R. S.
and Belagiannis, Vasileios
and Papa, Jo{\~a}o Paulo
and Nascimento, Jacinto C.
and Loog, Marco
and Lu, Zhi
and Cardoso, Jaime S.
and Cornebise, Julien",
title="The Importance of Skip Connections in Biomedical Image Segmentation",
booktitle="Deep Learning and Data Labeling for Medical Applications",
year="2016",
publisher="Springer International Publishing",
address="Cham",
pages="179--187",
isbn="978-3-319-46976-8"}

@article{oktay2018attention,
  title={Attention u-net: Learning where to look for the pancreas},
  author={Oktay, Ozan and Schlemper, Jo and Folgoc, Loic Le and Lee, Matthew and Heinrich, Mattias and Misawa, Kazunari and Mori, Kensaku and McDonagh, Steven and Hammerla, Nils Y and Kainz, Bernhard and others},
  journal={arXiv preprint arXiv:1804.03999},
  year={2018}}

@inproceedings{myronenko20193d,
  title={3D MRI brain tumor segmentation using autoencoder regularization},
  author={Myronenko, Andriy},
  booktitle={Brainlesion: Glioma, Multiple Sclerosis, Stroke and Traumatic Brain Injuries: 4th International Workshop, BrainLes 2018, Held in Conjunction with MICCAI 2018, Granada, Spain, September 16, 2018, Revised Selected Papers, Part II 4},
  pages={311--320},
  year={2019},
  organization={Springer}}

@article{lin2017focal,
  title={Focal Loss for Dense Object Detection},
  author={Lin, T},
  journal={arXiv preprint arXiv:1708.02002},
  year={2017}}

@InProceedings{10.1007/978-3-031-09002-8_9,
author="Nguyen-Truong, Hai
and Pham, Quan-Dung",
editor="Crimi, Alessandro
and Bakas, Spyridon",
title="Dice Focal Loss with ResNet-like Encoder-Decoder Architecture in 3D Brain Tumor Segmentation",
booktitle="Brainlesion: Glioma, Multiple Sclerosis, Stroke and Traumatic Brain Injuries",
year="2022",
publisher="Springer International Publishing",
address="Cham",
pages="97--105",
isbn="978-3-031-09002-8"}

@inproceedings{milletari2016v,
  title={V-net: Fully convolutional neural networks for volumetric medical image segmentation},
  author={Milletari, Fausto and Navab, Nassir and Ahmadi, Seyed-Ahmad},
  booktitle={2016 fourth international conference on 3D vision (3DV)},
  pages={565--571},
  year={2016},
  organization={Ieee}}

@inproceedings{li2019gradient,
  title={Gradient harmonized single-stage detector},
  author={Li, Buyu and Liu, Yu and Wang, Xiaogang},
  booktitle={Proceedings of the AAAI conference on artificial intelligence},
  volume={33},
  pages={8577--8584},
  year={2019}}

@article{ahamed2023comprehensive,
      title={Comprehensive framework for evaluation of deep neural networks in detection and quantification of lymphoma from PET/CT images: clinical insights, pitfalls, and observer agreement analyses}, 
      author={Shadab Ahamed and Yixi Xu and Sara Kurkowska and Claire Gowdy and Joo H. O and Ingrid Bloise and Don Wilson and Patrick Martineau and François Bénard and Fereshteh Yousefirizi and Rahul Dodhia and Juan M. Lavista and William B. Weeks and Carlos F. Uribe and Arman Rahmim},
      year={2024},
      eprint={2311.09614},
      archivePrefix={arXiv},
      primaryClass={cs.CV}, 
}

@article{farabet2012learning,
  title={Learning hierarchical features for scene labeling},
  author={Farabet, Clement and Couprie, Camille and Najman, Laurent and LeCun, Yann},
  journal={IEEE transactions on pattern analysis and machine intelligence},
  volume={35},
  number={8},
  pages={1915--1929},
  year={2012},
  publisher={IEEE}}

@article{chen2014semantic,
  title={Semantic image segmentation with deep convolutional nets and fully connected CRFs},
  author={Chen, Liang-Chieh},
  journal={arXiv preprint arXiv:1412.7062},
  year={2014}}

@ARTICLE{7478072,
  author={Shelhamer, Evan and Long, Jonathan and Darrell, Trevor},
  journal={IEEE Transactions on Pattern Analysis and Machine Intelligence}, 
  title={Fully Convolutional Networks for Semantic Segmentation}, 
  year={2017},
  volume={39},
  number={4},
  pages={640-651},
  doi={10.1109/TPAMI.2016.2572683}}

@article{XU2023106882,
title = {Automatic segmentation of prostate cancer metastases in PSMA PET/CT images using deep neural networks with weighted batch-wise dice loss},
journal = {Computers in Biology and Medicine},
volume = {158},
pages = {106882},
year = {2023},
issn = {0010-4825},
doi = {https://doi.org/10.1016/j.compbiomed.2023.106882},
url = {https://www.sciencedirect.com/science/article/pii/S0010482523003475},
author = {Yixi Xu and Ivan Klyuzhin and Sara Harsini and Anthony Ortiz and Shun Zhang and François Bénard and Rahul Dodhia and Carlos F. Uribe and Arman Rahmim and Juan {Lavista Ferres}}}

@article{harsini2023outcome,
  title={Outcome of patients with biochemical recurrence of prostate cancer after PSMA PET/CT-directed radiotherapy or surgery without systemic therapy},
  author={Harsini, Sara and Wilson, Don and Saprunoff, Heather and Allan, Hayley and Gleave, Martin and Goldenberg, Larry and Chi, Kim N and Kim-Sing, Charmaine and Tyldesley, Scott and B{\'e}nard, Fran{\c{c}}ois},
  journal={Cancer Imaging},
  volume={23},
  number={1},
  pages={27},
  year={2023},
  publisher={Springer}
}

@misc{cardoso2022monaiopensourceframeworkdeep,
      title={MONAI: An open-source framework for deep learning in healthcare}, 
      author={M. Jorge Cardoso and Wenqi Li and Richard Brown and Nic Ma and Eric Kerfoot and Yiheng Wang and Benjamin Murrey and Andriy Myronenko and Can Zhao and Dong Yang and Vishwesh Nath and Yufan He and Ziyue Xu and Ali Hatamizadeh and Andriy Myronenko and Wentao Zhu and Yun Liu and Mingxin Zheng and Yucheng Tang and Isaac Yang and Michael Zephyr and Behrooz Hashemian and Sachidanand Alle and Mohammad Zalbagi Darestani and Charlie Budd and Marc Modat and Tom Vercauteren and Guotai Wang and Yiwen Li and Yipeng Hu and Yunguan Fu and Benjamin Gorman and Hans Johnson and Brad Genereaux and Barbaros S. Erdal and Vikash Gupta and Andres Diaz-Pinto and Andre Dourson and Lena Maier-Hein and Paul F. Jaeger and Michael Baumgartner and Jayashree Kalpathy-Cramer and Mona Flores and Justin Kirby and Lee A. D. Cooper and Holger R. Roth and Daguang Xu and David Bericat and Ralf Floca and S. Kevin Zhou and Haris Shuaib and Keyvan Farahani and Klaus H. Maier-Hein and Stephen Aylward and Prerna Dogra and Sebastien Ourselin and Andrew Feng},
      year={2022},
      eprint={2211.02701},
      archivePrefix={arXiv},
      primaryClass={cs.LG},
      url={https://arxiv.org/abs/2211.02701}, 
}

@article{YEUNG2022102026,
title = {Unified Focal loss: Generalising Dice and cross entropy-based losses to handle class imbalanced medical image segmentation},
journal = {Computerized Medical Imaging and Graphics},
volume = {95},
pages = {102026},
year = {2022},
issn = {0895-6111},
doi = {https://doi.org/10.1016/j.compmedimag.2021.102026},
url = {https://www.sciencedirect.com/science/article/pii/S0895611121001750},
author = {Michael Yeung and Evis Sala and Carola-Bibiane Schönlieb and Leonardo Rundo},
}

@article{zhu2019anatomynet,
  title={AnatomyNet: deep learning for fast and fully automated whole-volume segmentation of head and neck anatomy},
  author={Zhu, Wentao and Huang, Yufang and Zeng, Liang and Chen, Xuming and Liu, Yong and Qian, Zhen and Du, Nan and Fan, Wei and Xie, Xiaohui},
  journal={Medical physics},
  volume={46},
  number={2},
  pages={576--589},
  year={2019},
  publisher={Wiley Online Library}
}

@InProceedings{10.1007/978-3-030-32226-7_10,
author="Liu, Qin
and Tang, Xiongfeng
and Guo, Deming
and Qin, Yanguo
and Jia, Pengfei
and Zhan, Yiqiang
and Zhou, Xiang
and Wu, Dijia",
editor="Shen, Dinggang
and Liu, Tianming
and Peters, Terry M.
and Staib, Lawrence H.
and Essert, Caroline
and Zhou, Sean
and Yap, Pew-Thian
and Khan, Ali",
title="Multi-class Gradient Harmonized Dice Loss with Application to Knee MR Image Segmentation",
booktitle="Medical Image Computing and Computer Assisted Intervention -- MICCAI 2019",
year="2019",
publisher="Springer International Publishing",
address="Cham",
pages="86--94",
isbn="978-3-030-32226-7"
}

@article{macmanus2009use,
  title={Use of PET and PET/CT for radiation therapy planning: IAEA expert report 2006--2007},
  author={MacManus, Michael and Nestle, Ursula and Rosenzweig, Kenneth E and Carrio, Ignasi and Messa, Cristina and Belohlavek, Otakar and Danna, Massimo and Inoue, Tomio and Deniaud-Alexandre, Elizabeth and Schipani, Stefano and others},
  journal={Radiotherapy and oncology},
  volume={91},
  number={1},
  pages={85--94},
  year={2009},
  publisher={Elsevier}
}

@article{hofheinz2013automatic,
  title={An automatic method for accurate volume delineation of heterogeneous tumors in PET},
  author={Hofheinz, Frank and Langner, Jens and Petr, Jan and Beuthien-Baumann, Bettina and Steinbach, J{\"o}rg and Kotzerke, J{\"o}rg and van den Hoff, J},
  journal={Medical physics},
  volume={40},
  number={8},
  pages={082503},
  year={2013},
  publisher={Wiley Online Library}
}

@article{fendler201768,
  title={68 Ga-PSMA PET/CT: Joint EANM and SNMMI procedure guideline for prostate cancer imaging: version 1.0},
  author={Fendler, Wolfgang P and Eiber, Matthias and Beheshti, Mohsen and Bomanji, Jamshed and Ceci, Francesco and Cho, Steven and Giesel, Frederik and Haberkorn, Uwe and Hope, Thomas A and Kopka, Klaus and others},
  journal={European journal of nuclear medicine and molecular imaging},
  volume={44},
  pages={1014--1024},
  year={2017},
  publisher={Springer}
}

@article{lauri2022psma,
  title={PSMA expression in solid tumors beyond the prostate gland: ready for theranostic applications?},
  author={Lauri, Chiara and Chiurchioni, Lorenzo and Russo, Vincenzo Marcello and Zannini, Luca and Signore, Alberto},
  journal={Journal of Clinical Medicine},
  volume={11},
  number={21},
  pages={6590},
  year={2022},
  publisher={MDPI}
}

@article{foster2014review,
  title={A review on segmentation of positron emission tomography images},
  author={Foster, Brent and Bagci, Ulas and Mansoor, Awais and Xu, Ziyue and Mollura, Daniel J},
  journal={Computers in biology and medicine},
  volume={50},
  pages={76--96},
  year={2014},
  publisher={Elsevier}
}

@article{gafita2019qpsma,
  title={qPSMA: semiautomatic software for whole-body tumor burden assessment in prostate cancer using 68Ga-PSMA11 PET/CT},
  author={Gafita, Andrei and Bieth, Marie and Kr{\"o}nke, Markus and Tetteh, Giles and Navarro, Fernando and Wang, Hui and G{\"u}nther, Elisabeth and Menze, Bjoern and Weber, Wolfgang A and Eiber, Matthias},
  journal={Journal of Nuclear Medicine},
  volume={60},
  number={9},
  pages={1277--1283},
  year={2019},
  publisher={Soc Nuclear Med}
}

@article{li2024automated,
  title={An Automated Deep Learning-Based Framework for Uptake Segmentation and Classification on PSMA PET/CT Imaging of Patients with Prostate Cancer},
  author={Li, Yang and Imami, Maliha R and Zhao, Linmei and Amindarolzarbi, Alireza and Mena, Esther and Leal, Jeffrey and Chen, Junyu and Gafita, Andrei and Voter, Andrew F and Li, Xin and others},
  journal={Journal of Imaging Informatics in Medicine},
  pages={1--10},
  year={2024},
  publisher={Springer}
}

@article{Zhang2021DiceLoss,
  author    = {Yue Zhang and Shijie Liu and Chunlai Li and Jianyu Wang},
  title     = {Rethinking the Dice Loss for Deep Learning Lesion Segmentation in Medical Images},
  journal   = {Journal of Shanghai Jiaotong University (Science)},
  volume    = {26},
  number    = {1},
  pages     = {93--102},
  year      = {2021},
  doi       = {10.1007/s12204-021-2264-x},
  url       = {https://doi.org/10.1007/s12204-021-2264-x},
  issn      = {1995-8188}
}

@inproceedings{Kofler2023BlobLoss,
  title={Blob loss: Instance imbalance aware loss functions for semantic segmentation},
  author={Kofler, Florian and Shit, Suprosanna and Ezhov, Ivan and Fidon, Lucas and Horvath, Izabela and Al-Maskari, Rami and Li, Hongwei Bran and Bhatia, Harsharan and Loehr, Timo and Piraud, Marie and others},
  booktitle={International Conference on Information Processing in Medical Imaging},
  pages={755--767},
  year={2023},
  organization={Springer}
}

@article{bhandary2024segmentation,
  title={Segmentation of Prostate Tumour Volumes from PET Images is a Different Ball Game},
  author={Bhandary, Shrajan and Kuhn, Dejan and Babaiee, Zahra and Fechter, Tobias and Spohn, Simon KB and Zamboglou, Constantinos and Grosu, Anca-Ligia and Grosu, Radu},
  journal={arXiv preprint arXiv:2407.10537},
  year={2024}
}

@inproceedings{toosi2024segment,
  title={How to Segment in 3D Using 2D Models: Automated 3D Segmentation of Prostate Cancer Metastatic Lesions on PET Volumes Using Multi-angle Maximum Intensity Projections and Diffusion Models},
  author={Toosi, Amirhosein and Harsini, Sara and B{\'e}nard, Fran{\c{c}}ois and Uribe, Carlos and Rahmim, Arman},
  booktitle={MICCAI Workshop on Deep Generative Models},
  pages={212--221},
  year={2024},
  organization={Springer}
}

@article{constantino2025use,
  title={The use of maximum-intensity projections and deep learning adds value to the fully automatic segmentation of lesions avid for [18f] fdg and [68ga] ga-psma in pet/ct},
  author={Constantino, Cl{\'a}udia S and Oliveira, Francisco PM and Machado, Marisa and Vinga, Susana and Costa, Durval C},
  journal={Journal of Nuclear Medicine},
  volume={66},
  number={5},
  pages={795--801},
  year={2025},
  publisher={Society of Nuclear Medicine}
}

@book{levin2017markov,
  title={Markov chains and mixing times},
  author={Levin, David A and Peres, Yuval},
  volume={107},
  year={2017},
  publisher={American Mathematical Soc.}
}

@article{peyre2019computational,
  title={Computational optimal transport: With applications to data science},
  author={Peyr{\'e}, Gabriel and Cuturi, Marco and others},
  journal={Foundations and Trends{\textregistered} in Machine Learning},
  volume={11},
  number={5-6},
  pages={355--607},
  year={2019},
  publisher={Now Publishers, Inc.}
}

@inproceedings{wang2024sam,
  title={Sam-med3d: towards general-purpose segmentation models for volumetric medical images},
  author={Wang, Haoyu and Guo, Sizheng and Ye, Jin and Deng, Zhongying and Cheng, Junlong and Li, Tianbin and Chen, Jianpin and Su, Yanzhou and Huang, Ziyan and Shen, Yiqing and others},
  booktitle={European Conference on Computer Vision},
  pages={51--67},
  year={2024},
  organization={Springer}
}

@inproceedings{cciccek20163d,
  title={3D U-Net: learning dense volumetric segmentation from sparse annotation},
  author={{\c{C}}i{\c{c}}ek, {\"O}zg{\"u}n and Abdulkadir, Ahmed and Lienkamp, Soeren S and Brox, Thomas and Ronneberger, Olaf},
  booktitle={International conference on medical image computing and computer-assisted intervention},
  pages={424--432},
  year={2016},
  organization={Springer}
}

@inproceedings{hatamizadeh2022unetr,
  title={Unetr: Transformers for 3d medical image segmentation},
  author={Hatamizadeh, Ali and Tang, Yucheng and Nath, Vishwesh and Yang, Dong and Myronenko, Andriy and Landman, Bennett and Roth, Holger R and Xu, Daguang},
  booktitle={Proceedings of the IEEE/CVF winter conference on applications of computer vision},
  pages={574--584},
  year={2022}
}

@inproceedings{lee2022towards,
  title={Towards flexible inductive bias via progressive reparameterization scheduling},
  author={Lee, Yunsung and Lee, Gyuseong and Ryoo, Kwangrok and Go, Hyojun and Park, Jihye and Kim, Seungryong},
  booktitle={European Conference on Computer Vision},
  pages={706--720},
  year={2022},
  organization={Springer}
}

@article{lu2022bridging,
  title={Bridging the gap between vision transformers and convolutional neural networks on small datasets},
  author={Lu, Zhiying and Xie, Hongtao and Liu, Chuanbin and Zhang, Yongdong},
  journal={Advances in Neural Information Processing Systems},
  volume={35},
  pages={14663--14677},
  year={2022}
}

@article{jafari2024convolutional,
  title={A convolutional neural network--based system for fully automatic segmentation of whole-body [68Ga] Ga-PSMA PET images in prostate cancer},
  author={Jafari, Esmail and Zarei, Amin and Dadgar, Habibollah and Keshavarz, Ahmad and Manafi-Farid, Reyhaneh and Rostami, Habib and Assadi, Majid},
  journal={European journal of nuclear medicine and molecular imaging},
  volume={51},
  number={5},
  pages={1476--1487},
  year={2024},
  publisher={Springer}
}

@article{kendrick2022fully,
  title={Fully automatic prognostic biomarker extraction from metastatic prostate lesion segmentations in whole-body [68Ga] Ga-PSMA-11 PET/CT images},
  author={Kendrick, Jake and Francis, Roslyn J and Hassan, Ghulam Mubashar and Rowshanfarzad, Pejman and Ong, Jeremy SL and Ebert, Martin A},
  journal={European Journal of Nuclear Medicine and Molecular Imaging},
  volume={50},
  number={1},
  pages={67--79},
  year={2022},
  publisher={Springer}
}

@article{zhao2020deep,
  title={Deep neural network for automatic characterization of lesions on 68Ga-PSMA-11 PET/CT},
  author={Zhao, Yu and Gafita, Andrei and Vollnberg, Bernd and Tetteh, Giles and Haupt, Fabian and Afshar-Oromieh, Ali and Menze, Bjoern and Eiber, Matthias and Rominger, Axel and Shi, Kuangyu},
  journal={European journal of nuclear medicine and molecular imaging},
  volume={47},
  number={3},
  pages={603--613},
  year={2020},
  publisher={Springer}
}

@inproceedings{sudre2017generalised,
  title={Generalised dice overlap as a deep learning loss function for highly unbalanced segmentations},
  author={Sudre, Carole H and Li, Wenqi and Vercauteren, Tom and Ourselin, Sebastien and Jorge Cardoso, M},
  booktitle={International Workshop on Deep Learning in Medical Image Analysis},
  pages={240--248},
  year={2017},
  organization={Springer}
}

@article{taghanaki2019combo,
  title={Combo loss: Handling input and output imbalance in multi-organ segmentation},
  author={Taghanaki, Saeid Asgari and Zheng, Yefeng and Zhou, S Kevin and Georgescu, Bogdan and Sharma, Puneet and Xu, Daguang and Comaniciu, Dorin and Hamarneh, Ghassan},
  journal={Computerized Medical Imaging and Graphics},
  volume={75},
  pages={24--33},
  year={2019},
  publisher={Elsevier}
}

@article{wu2016bridging,
  title={Bridging category-level and instance-level semantic image segmentation},
  author={Wu, Zifeng and Shen, Chunhua and Hengel, Anton van den},
  journal={arXiv preprint arXiv:1605.06885},
  year={2016}
}

@article{begoli2019need,
  title={The need for uncertainty quantification in machine-assisted medical decision making},
  author={Begoli, Edmon and Bhattacharya, Tanmoy and Kusnezov, Dimitri},
  journal={Nature Machine Intelligence},
  volume={1},
  number={1},
  pages={20--23},
  year={2019},
  publisher={Nature Publishing Group UK London}
}

\appendix
\section{Training Behavior Across Networks and Losses}
\label{ap:train_curve}

\begin{figure}[tb!]
   \centering
   \includegraphics[width=\linewidth, keepaspectratio]{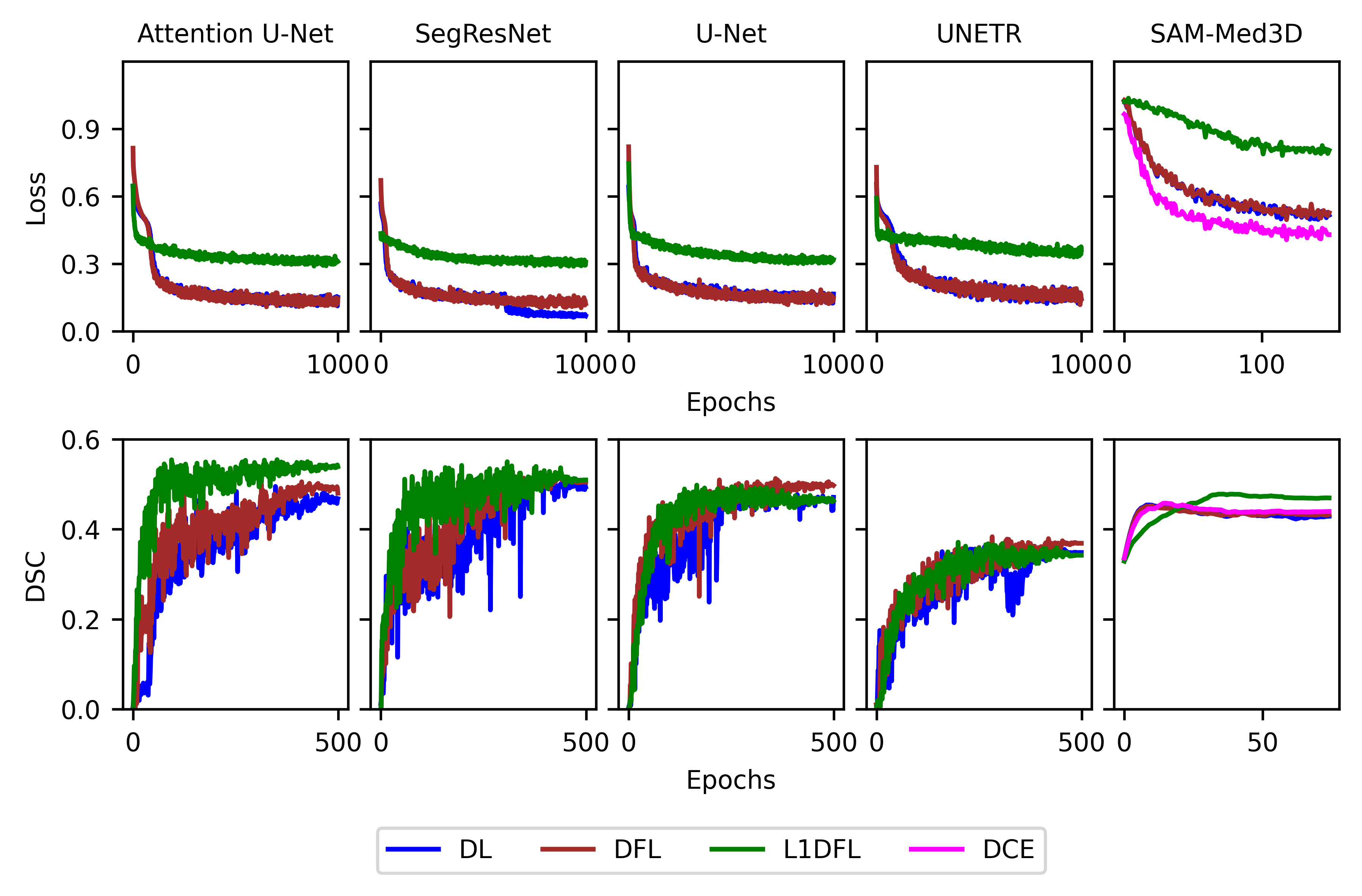}
   \caption{Training and validation curves across five segmentation networks and four loss functions, DL, DFL, L1DFL, and DCE (the original loss function for SAM-Med3D). The top row shows training loss trajectories, while the bottom row shows validation Dice Similarity Coefficient evolution over epochs. Columns correspond to different network architectures and loss functions are color coded. SAM-Med3D training loss values were rescaled by a factor of 10 for visualization.
   \label{fig:train}
    } 
\end{figure}

Figure \ref{fig:train} presents the training loss (top row) and validation DSC (bottom row) curves across networks and loss functions. Across all models, the training losses decreased consistently, confirming stable convergence. Attention U-Net, SegResNet, and U-Net exhibited rapid early loss reduction, with DL and DFL converging toward the lowest final loss values. In contrast, L1DFL maintained a higher residual loss but plateaued steadily, suggesting a different optimization dynamic. UNETR followed a similar pattern, albeit with slower convergence. SAM-Med3D showed higher apparent loss values.

Validation DSC curves highlighted differences in generalization. L1DFL consistently yielded the highest DSC values across all networks, reaching approximately $0.55$ – $0.58$, and showing stable trajectories with minimal late-epoch decline. DFL provided intermediate performance, while DL typically converged more slowly and plateaued at lower DSC values. Interestingly, U-Net and Attention U-Net showed sharp early gains under L1DFL, while UNETR’s DSC curves lagged behind, indicating slower but eventually comparable improvements. For SAM-Med3D, DSC curves converged more quickly and stabilized by 20 epochs, reflecting its pretraining advantage.

\section{Model Evaluation}
\label{ap:model_evaluation}

We assessed the performance of the loss functions on the overall test data. We made predictions on the PET/CT (only PET for SAM-Med3D)  whole-body images using the sliding-window technique \cite{cardoso2022monaiopensourceframeworkdeep} with a window of dimension (128, 128, 128) for all networks. The test set predictions were resampled to the coordinates of the original ground truth masks to compute the evaluation metrics. In addition to assessing performance on the overall test set, we performed evaluations based on two different lesion scenarios, single and multiple lesion scenarios. First, we categorized the set of images (\( I \)) into two subsets: images with a single-lesion (\( S \)) and images with multiple-lesion (\( M \)). For an image \( x \) with \( n(G_l) \) ground truth lesions, we defined these groups as follows:

\begin{equation*}
S = \{ x \in I \mid n(G_l) = 1 \}
\end{equation*}

\begin{equation*}
M = \{ x \in I \mid n(G_l) \geq 2 \}
\end{equation*}
Then, for each scenario, as well as on the overall test set, we performed patient-level and lesion-level assessments based on the methodologies below.

\subsubsection{Patient-Level Analysis}
\paragraph{Segmentation Metrics:}
We evaluated the segmentation performance of the loss functions using the DSC. We report the mean DSC together with the standard deviation and the median DSC with the inter-quartile ranges. Let \( G \) and \( P \) represent the ground truth and predicted masks of an image at the patient level. The patient-level DSC is defined as:

\begin{equation}
\text{DSC} = \frac{2 |G \cap P|}{|G| + |P|}
\end{equation}

\begin{figure}[ht]
   \begin{centering}
   \includegraphics[width=12cm]{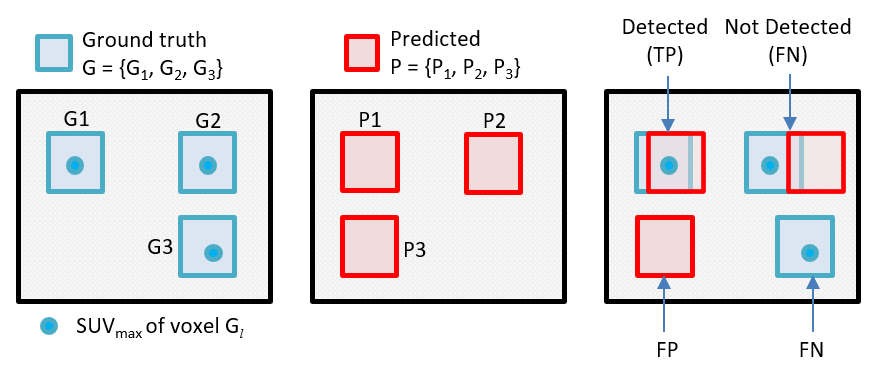}
   \caption{Illustration for defining a true positive detection based on an overlap with the voxel containing the maximum standardized uptake value (SUVmax) in the ground truth lesion. G is the set of ground truth lesions and P is the set of predicted lesions.
   \label{fig:detection} 
    } 
    \end{centering}
\end{figure}

\paragraph{Detection Metrics:}
We defined true positive (TP), false positive (FP), and false negative (FN) detections at the patient level, reporting their mean values across the test set and computed F1 scores per patient. A detection is considered as TP if there is a matched pair of \( G_l \) and \( P_l \) such that $G_{l,\text{SUVmax}} \cap P_l \neq 0$, where \( G_{l,\text{SUVmax}} \) is the voxel in \( G_l \) with the maximum standardized uptake value (SUVmax) \cite{ahamed2023comprehensive}. For a prediction \( P_l \), for which there is no corresponding overlap with a ground truth lesion \( G_l \), the detection is designated as an FP. Similarly, a false negative occurs when $G_{l,\text{SUVmax}} \cap P_l = 0$. We provide a visual illustration of the definition of the detection metrics in Figure \ref{fig:detection}. The F1 score is thus computed as:

\begin{equation}
\text{F1} = \frac{\text{TP}}{\text{TP} + \frac{1}{2} (\text{FP} + \text{FN})}
\end{equation}

\paragraph{Clinical Metrics:}
For each scenario, we analyzed the performance of the loss functions across different groupings of molecular tumor volume. Specifically, we evaluated the performance of the loss functions on DSC on different thresholds of total molecular tumor volume (TMTV). We computed thresholds $(t)$ based on the inter-quartile range from 0 to the 85th percentile of the ground truth TMTV values and calculated median DSC for all lesions $(l)$ where volume $(v_l > t)$. Additionally, for the multiple-lesion scenario only, we assessed the loss function performance based on the spatial extent of lesion distribution, $D_{\text{max}}$, which is measured as the maximum distance between any pair of foreground voxels in the image. 

For a given ground truth mask, let \(\mathbf{v}_i\) denote foreground voxels, where \(i = 1, 2, \ldots, N\) and \(N\) is the total number of lesion voxels. We calculated the Euclidean distance between every pair of lesion voxels, \( \mathbf{v}_i \) and \( \mathbf{v}_j \), accounting for voxel spacing in each dimension. For voxel coordinates \((x_i, y_i, z_i)\) and \((x_j, y_j, z_j)\) with spacing \((s_x, s_y, s_z)\), the distance \( d_{ij} \) is given by equation (\ref{eq:Dmax}) below:

   \begin{equation}
   \label{eq:Dmax}
   d_{ij} = \sqrt{ \left( s_x (x_i - x_j) \right)^2 + \left( s_y (y_i - y_j) \right)^2 + \left( s_z (z_i - z_j) \right)^2 }
   \end{equation}
The lesion dissemination, \(D_{\text{max}} \), is calculated as the maximum distance among all calculated distances; \(D_{\text{max}} = \max_{i, j} \, d_{ij}\). We categorized the calculated distances into inter-quartile ranges (IQR) indicating the first, second, third and fourth IQR and analyzed the performance of the loss functions on each group. 

\subsubsection{Lesion-Level Analysis}
\paragraph{Segmentation Metrics:}
For lesion-level analysis, we defined the DSC based on the overlap between each predicted lesion mask and its corresponding ground truth mask. We identified which predicted lesion voxels corresponded to the ground truth voxels, especially in the scenario of multiple lesions in an image, based on a voxel-wise matching strategy. Specifically, lesions were segmented as individual connected components in both the ground truth and predicted masks using 18-connectivity to ensure spatial continuity within each identified lesion. This produced unique integer labels for each lesion. Each connected component in the predicted mask was then assessed for spatial overlap with each connected component in the ground truth mask. 

For each pair of ground truth and predicted lesions, a match was defined if there was any voxel overlap between the components (i.e., if any voxels in a predicted lesion occupied the same spatial locations as those in a ground truth lesion). This was computed by checking if any intersecting voxels existed between the two labeled components. When an overlap was identified, the corresponding pair of ground truth and predicted lesion labels was recorded as a matched lesion pair. These matches were used to compute lesion-wise metrics, including lesion-level DSC, by comparing the voxel distributions in each matched pair of lesions. We defined the lesion-wise DSC as:

\begin{equation}
\text{DSC}_l = \frac{2 |G_l \cap P_l|}{|G_l| + |P_l|}
\end{equation}
where \( G_l \) and \( P_l \) denote the ground truth and predicted masks of a specific lesion. For a given ground truth lesion with no match, $|G_l \cap P_l| = 0$ yielding a DSC of $0$. 

\paragraph{Clinical Metrics:}
At the lesion level, analysis was performed only on the molecular tumor volumes of the lesions. Similar to the threshold analysis performed at the patient level, we calculated thresholds $(t)$ based on the inter-quartile range from 0 to the 85th percentile of the ground truth MTV values (molecular volume of individual lesions) and calculated median DSC for all lesions $(l)$ where volume $(v_l > t)$. For each scenario, we analyzed the performance of the loss functions across different groupings of MTV.

\end{document}